\documentclass[twocolumn,iop,apj]{emulateapj} 
\usepackage{times}
\usepackage{natbib}

\extrafloats{100}
\usepackage[citecolor=blue,hypertexnames=false]{hyperref}
\hypersetup{
    colorlinks=true,       
    linkcolor=blue,                 
    filecolor=magenta,      
    urlcolor=blue
}

\usepackage{verbatim}
\usepackage[normalem]{ulem}
\usepackage[usenames, dvipsnames]{color}
\usepackage[section]{placeins}

\graphicspath{{images/}}
\newcommand{\myemail}{Corresponding Mail}

\newcommand{\fermilat}{\emph{Fermi}-LAT}

\newcommand{\galprop}{GALPROP}
\newcommand{\gray}{$\gamma$-ray}

\newcommand{\helmod}{\textsc{HelMod}}

\slugcomment{Draft: \today}

\shorttitle{Local interstellar spectra of secondary nuclei}
\shortauthors{Boschini et al.}

\usepackage{amsmath} 

\begin{document}


\title{


Deciphering the local Interstellar spectra of secondary nuclei with \galprop{}/\helmod{} framework and a hint for primary lithium in cosmic rays
 
}


\author{
M.~J.~Boschini\altaffilmark{1,2},  
S.~{Della~Torre}\altaffilmark{1}, 
M.~Gervasi\altaffilmark{1,3}, 
D.~Grandi\altaffilmark{1,3},
G.~J\'{o}hannesson\altaffilmark{4,5}, 
G.~{La~Vacca}\altaffilmark{1,3}, 
N.~Masi\altaffilmark{6},
I.~V.~Moskalenko\altaffilmark{7,8}, 
S.~Pensotti\altaffilmark{1,3}, 
T.~A.~Porter\altaffilmark{7,8}, 
L.~Quadrani\altaffilmark{6,9}, 
P.~G.~Rancoita\altaffilmark{1},
D.~Rozza\altaffilmark{1,3}, 
and M.~Tacconi\altaffilmark{1,3}
}
\email{\myemail}


\altaffiltext{1}{INFN, Milano-Bicocca, Milano, Italy}
\altaffiltext{2}{also CINECA, Segrate, Milano, Italy}
\altaffiltext{3}{also Physics Department, University of Milano-Bicocca, Milano, Italy}
\altaffiltext{4}{Science Institute, University of Iceland, Dunhaga 3, IS-107 Reykjavik, Iceland}
\altaffiltext{5}{also NORDITA,  Roslagstullsbacken 23, 106 91 Stockholm, Sweden}
\altaffiltext{6}{INFN, Bologna, Italy}
\altaffiltext{7}{Hansen Experimental Physics Laboratory, Stanford University, Stanford, CA 94305}
\altaffiltext{8}{Kavli Institute for Particle Astrophysics and Cosmology, Stanford University, Stanford, CA 94305}
\altaffiltext{9}{also, Physics Department, University of Bologna, Bologna, Italy}


\begin{abstract}

Local interstellar spectra (LIS) of secondary cosmic ray (CR) nuclei, lithium, beryllium, boron, and partially secondary nitrogen, are derived in the rigidity range from 10 MV to $\sim$200 TV using the most recent experimental results combined with the state-of-the-art models for CR propagation in the Galaxy and in the heliosphere. The lithium spectrum appears somewhat flatter at high energies compared to other secondary species that may imply a primary lithium component. Two propagation packages, \galprop{} and \helmod{}, are combined to provide a single framework that is run to reproduce direct measurements of CR species at different modulation levels, and at both polarities of the solar magnetic field. An iterative maximum-likelihood method is developed that uses \galprop-predicted LIS as input to \helmod{}, which provides the modulated spectra for specific time periods of the selected experiments for the model-data comparison. The proposed LIS accommodate the low-energy interstellar spectra measured by Voyager 1, HEAO-3, and ACE/CRIS as well as the high-energy observations by PAMELA, AMS-02, and earlier experiments that are made deep in the heliosphere. The interstellar and heliospheric propagation parameters derived in this study are consistent with our earlier results for propagation of CR protons, helium, carbon, oxygen, antiprotons, and electrons.
\end{abstract}



\section{Introduction} \label{Intro}

Cosmic rays (CRs) are the only pieces of matter available to us that come from large Galactic and extragalactic distances. Their spectrum and composition provide invaluable information about their origin and propagation history. The bulk of Galactic CRs is associated with the most energetic events such as supernova explosions, but some fraction may also come from pulsars, interstellar shocks, neutron star mergers, and perhaps from even more exotic and less studied processes. The origin of extragalactic CRs is still a mystery with speculations ranging from nuclei of active galaxies to gamma-ray bursts and primordial shocks.

Virtually all hydrogen $^1$H and most of helium $^4$He were produced in the first minutes after the Big Bang along with the trace amounts of deuterium $^2$H, light helium isotope $^3$He, and $^7$Li. Almost all other varieties of nuclei species are the result of the stellar thermonuclear burning and explosive nucleosynthesis. This material is then mixed into the interstellar medium, fragmented and reprocessed by energetic CR particles interacting with the interstellar gas. The new shocks propagating through such a mixture accelerate all available species producing fresh CRs that are mixing up with older CRs produced by supernova remnants, which went off millions of years ago when the human race did not even exist yet. 

Although, fragmentation of CR species does not play a major role in chemical evolution of the Galaxy, it is the main source of isotopes that are not synthesized in stars or depleted in the process of thermonuclear burning. The nuclides of this kind are isotopes of Li, Be, and B, that are called secondary. Fragmentation of CR nuclei produces also all other isotopes that makes the composition of CRs noticeably different from the composition of their sources and the interstellar matter. For instance, the amounts of primary and secondary nitrogen in CRs are about equal. The relative amounts and spectra of those so-called secondary isotopes reflect the processes of their production as well as their acceleration and propagation history. Therefore, studies of CR composition and spectra are able to shed light on the global properties and history of our Galaxy, CR sources and acceleration processes, and properties of the components of the interstellar medium (e.g., interstellar gas distribution, spectrum of turbulence etc.). 

The observed abundances of stable secondary CR nuclei (e.g., $^3$He, Li, Be, B, Sc, Ti, V) and radioactive isotopes with half-life of $\sim$1 Myr ($^{10}$Be, $^{26}$Al, $^{36}$Cl, $^{54}$Mn) allow the determination of the Galactic halo size and diffusion coefficient \citep{1998A&A...337..859P,1998ApJ...509..212S,2001AdSpR..27..717S,1998ApJ...506..335W,2001ICRC....5.1836M, 2011ApJ...729..106T,2016ApJ...824...16J}. K-capture isotopes in CRs (e.g., $^{49}$V, $^{51}$Cr) can be used to study energy-dependent effects \citep[][]{1998A&A...336L..61S,2001AdSpR..27..737J,2003JGRA..108.8033N}, such as the diffusive reacceleration, because their lifetime depends on the rates of electron attachment and stripping in the interstellar gas. Trans-iron CR nuclei abundances ($Z\ge29$) is invaluable source of information about our local environment. Their large fragmentation cross sections \citep[e.g., see Figure 1 in][]{1996PhRvC..54.1329W} imply that they originate in the local sources. 

Most of these measurements are done at low energies deep inside the heliosphere, where the solar modulation is significant. Meanwhile, many parameters of the interstellar propagation models are derived using these low-energy measurements, which are then extrapolated to the TeV energies and above. Such models, in turn, are used for interpretation of data of CR missions and space telescopes and to search for signatures of new physics. Therefore, the determination of the true local interstellar spectra (LIS) of secondary species, Li, Be, B, is of considerable interest for the astrophysics and particle physics communities. 

In this paper, we use an updated version of a 2D Monte Carlo code for heliospheric propagation \helmod{} \citep{2018AdSpR..62.2859B,2019HelMod} combined with the latest version 56 of the interstellar propagation code \galprop{} \citep{2017ApJ...846...67P,2018ApJ...856...45J} to take advantage of significant progress in CR measurements to derive the LIS for Li, Be, and B. The \helmod{} model solves the Parker equation \citep{1965P&SS...13....9P} and includes all relevant effects including a full description of the diffusion tensor. Our method and approach are described in a series of recent papers devoted to the LIS of CR $p$, $\bar{p}$, $e^-$, and He, C, O nuclei \citep{2017ApJ...840..115B,2018ApJ...854...94B,2018ApJ...858...61B}. 

\section{CR transport in the Galaxy and the heliosphere} \label{sec2}

Here we provide short descriptions of the two dedicated codes that are used in the present work and that complement each other: \galprop{}\footnote{Available from http://galprop.stanford.edu \label{galprop-site}} -- for description of the interstellar propagation, and \helmod{}\footnote{http://www.helmod.org/} -- for description of the heliospheric transport.  More details can be found in the referenced papers.

\subsection{\galprop{} Model for Galactic CR Propagation and diffuse emission}\label{galprop}

The state-of-the-art propagation code called \galprop{} is widely employed for modeling of CRs and associated emissions from the Milky Way, and now has about 23 years of development behind it \citep{1998ApJ...493..694M,1998ApJ...509..212S}. The \galprop{} code uses information from astronomy, particle, and nuclear physics to predict CRs, \gray{s}, synchrotron emission and its polarization in a self-consistent manner -- it provides the modeling framework unifying many results of individual measurements in physics and astronomy spanning in energy coverage, types of instrumentation, and the nature of detected species. The \galprop{} code range of physical validity covers sub-keV -- PeV energies for particles and from $10^{-6}$ eV -- PeV for photons. Over the years the project has been widely recognized as a standard modeling tool for Galactic CR propagation and associated diffuse emissions (radio, X-rays, \gray{s}). The \galprop{} code is \emph{public} and is extensively used by many experimental groups, and by 1000s of individual researchers worldwide for interpretation of their data and for making predictions. 

The key concept underlying the \galprop{} code is that various kinds of data, e.g., direct CR measurements, $\bar{p}$, $e^\pm$, \gray{s}, synchrotron radiation, and so forth, are all related to the same Galaxy and hence have to be modeled self-consistently \citep{1998A&A...338L..75M}. The goal for the \galprop{}-based models is to be as realistic as possible and to make use of available astronomical information, nuclear and particle data, with a minimum of simplifying assumptions \citep{2007ARNPS..57..285S}.

The \galprop{} code \citep{1998ApJ...509..212S} solves a system of about 90 transport equations (time-dependent partial differential equations in 3D or 4D: spatial variables plus energy) with a given source distribution and boundary conditions for all CR species ($^1$H-$^{64}$Ni, $\bar{p}$, $e^\pm$). This includes convection, distributed reacceleration, energy losses, nuclear fragmentation, radioactive decay, and production of secondary particles and isotopes. The numerical solution is based on a Crank-Nicholson implicit second-order scheme \citep{1992nrfa.book.....P}. The spatial boundary conditions assume free particle escape. For a given halo size the diffusion coefficient, as a function of momentum and propagation parameters, is determined from secondary-to-primary nuclei ratios, typically B/C, [Sc+Ti+V]/Fe, and/or $\bar{p}/p$. If reacceleration is included, the momentum-space diffusion coefficient $D_{pp}$ is related to the spatial coefficient $D_{xx}$ ($= \beta D_0 R^\delta$) \citep{1994ApJ...431..705S}, where $\delta = 1/3$ for a Kolmogorov spectrum of interstellar turbulence or $\delta = 1/2$ for an Iroshnikov-Kraichnan cascade, $R$ is the magnetic rigidity. 

The injection spectra of CR species are parametrized by the rigidity-dependent function:
\begin{equation}  \label{eq:1}
q(R) \propto (R/R_0)^{-\gamma_0}\prod_{i=0}^2\bigg[1 + (R/R_i)^\frac{\gamma_i - \gamma_{i+1}}{s_i}\bigg]^{s_i},
\end{equation}
where $\gamma_{i =0,1,2,3}$ are the spectral indices, $R_{i = 0,1,2}$ are the break rigidities, $s_i$ are the smoothing parameters ($s_i$ is negative/positive for $|\gamma_i |\lessgtr |\gamma_{i+1} |$). 

The \galprop{} code computes a complete network of primary, secondary, and tertiary CR production starting from input source abundances. \galprop{} includes K-capture, electron pick up and stripping processes \citep{1973RvMP...45..273P,1978PhDT........12W,1979PhDT........67C}, and knock-on electrons \citep{1966PhRv..150.1088A, 2003ApJ...594..709B}. Cross-sections are based on the extensive LANL database, nuclear codes, and parameterizations \citep{2004AdSpR..34.1288M}, see also a compilation in \citet{2018PhRvC..98c4611G}. The most important isotopic production cross-sections are calculated using our fits to major production channels \citep{2003ICRC....4.1969M, 2003ApJ...586.1050M, 2018PhRvC..98c4611G}. Other cross-sections are computed using phenomenological codes \citep{2003ApJS..144..153W, 1998ApJ...501..911S} renormalized to the data where they exists. The nuclear reaction network is built using the Nuclear Data Sheets. 

\galprop{} calculates production of secondary particles in $pp$-, $pA$-, $Ap$-, $AA$-interactions. Calculations of $\bar{p}$ production and propagation are detailed in \citet{2002ApJ...565..280M}, \citet{2003ApJ...586.1050M}, and \citet{2015ApJ...803...54K,2019arXiv190405129K}. Production of neutral mesons ($\pi^0$, $K^0$, $\bar{K}^0$, etc.), and secondary $e^\pm$ is calculated using the formalism by \citet{1986A&A...157..223D,1986ApJ...307...47D}, as described in \citet{1998ApJ...493..694M}, or recent parameterizations by \citet{2006ApJ...647..692K}, \citet{2012PhRvD..86d3004K}, \citet{2014ApJ...789..136K,2019arXiv190405129K}. \gray{} production and synchrotron emission are calculated using the propagated CR distributions, including primary $e^-$, secondary $e^\pm$, and for \gray{s} -- including secondary $p$ from inelastic processes \citep{2004ApJ...613..962S,2010ApJ...722L..58S,2008ApJ...682..400P, 2013MNRAS.436.2127O}. 

More details on \galprop{} including the description of all involved processes and reactions can be found in dedicated publications \citep{1997A&A...325..401M,1998ApJ...493..694M, 2000ApJ...528..357M,1998ApJ...509..212S,  2000ApJ...537..763S, 2004ApJ...613..962S, 2007ARNPS..57..285S, 2010ApJ...722L..58S,2011A&A...534A..54S, 2002ApJ...565..280M, 2003ApJ...586.1050M, PoS(ICRC2017)279, 2006ApJ...642..902P,2011CoPhC.182.1156V,2012ApJ...752...68V, 2013MNRAS.436.2127O, 2017ApJ...846...67P,2019arXiv190902223P, 2018ApJ...856...45J,2019ApJ...879...91J,2018PhRvC..98c4611G}.

\subsection{\helmod{} Model for heliospheric transport}\label{Sect::Helmod}

The combined effects of the intense solar wind and solar magnetic field modify the local interstellar space and develop a bubble-like structure surrounding the whole solar system that is called the heliosphere. The heliosphere affects the propagation of CR particles up to $\sim$50 GV in rigidity and requires a dedicated modeling to understand all factors involved \citep[see discussion in][]{2017ApJ...840..115B}. CR propagation in the heliosphere was first studied by \citet{1965P&SS...13....9P}, who formulated the transport equation also referred to as Parker equation~\citep[see, e.g., discussion in][and references therein]{Bobik2011ApJ}:
\begin{align}
\label{EQ::FPE}
 \frac{\partial U}{\partial t}= &\frac{\partial}{\partial x_i} \left( K^S_{ij}\frac{\partial \mathrm{U} }{\partial x_j}\right)\\
&+\frac{1}{3}\frac{\partial V_{ \mathrm{sw},i} }{\partial x_i} \frac{\partial }{\partial T}\left(\alpha_{\mathrm{rel} }T\mathrm{U} \right)
- \frac{\partial}{\partial x_i} [ (V_{ \mathrm{sw},i}+v_{d,i})\mathrm{U}],\nonumber
\end{align}
where $U$ is the number density of CR species per unit of kinetic energy $T$, $t$ is the time, $V_{ \mathrm{sw},i}$ is the solar wind velocity along the axis $x_i$, $K^S_{ij}$ is the symmetric part of the diffusion tensor, $v_{d,i}$ is the particle magnetic drift velocity (related to the antisymmetric part of the diffusion tensor), and finally $\alpha_{\mathrm{rel} }=\frac{T+2m_r c^2}{T+m_r c^2} $, with $m_r$ the particle rest mass in units of GeV/nucleon. Parker's transport equation describes: (i) the \textit{diffusion} of CR species due to magnetic irregularities, (ii) the so-called \textit{adiabatic-energy changes} associated with expansions and compressions of cosmic radiation, (iii) an \textit{effective convection} resulting from the convection with \textit{solar wind} (SW, with velocity $\vec{V}_{{\rm sw}}$), and (iv) the drift effects related to the \textit{drift velocity} ($\vec{v}_d$).

The influence of heliospheric propagation on the spectra of CR species is called the solar modulation. Its overall effect leads to the suppression of the low-energy part in the spectra of CR species, while the amplitude of the suppression depends on the solar activity, particle charge sign, polarity of the solar magnetic field and other conditions. In this work, the particle transport within the heliosphere, is treated by means of \helmod{} model \citep[and reference therein]{2019HelMod}. The \helmod{} model, now version 4.0, numerically solves the  \citet{1965P&SS...13....9P}  transport equation using a Monte Carlo approach involving stochastic differential equations \citep[see a discussion in, e.g.,][]{Bobik2011ApJ,BobikEtAl2016}. The particle transport within the heliosphere is computed from the outer boundary (i.e.\ the heliopause) down to Earth orbit. In this latest version the actual dimensions of the heliosphere and its boundaries were taken into account based on Voyager~1 measurements \citep{2019HelMod}. 

The heliopause (HP) represents the extreme limit beyond which solar modulation does not affect CR flux. Thus, the CR spectra measured by Voyager 1 outside HP are the truly pristine LIS of CR species\footnote{Voyager 2 is now in the interstellar space confirming the data from its sister spacecraft Voyager 1 \citep{2019Voyager2}.}. Using the Parker's model of the heliosphere \citep{1961ApJ...134...20P,1963idp..book.....P} in combination with Voyager~1 observations, we were able to estimate the time dependence of positions of the termination shock (TS, $R_{\rm TS}$) and the HP ($R_{\rm HP}$) as \citep{2019HelMod}:
\begin{align}
R_{\rm TS} &= R_{\rm obs} \left( \frac{\rho_{\rm obs} u_{\rm obs}^2}{P_{\rm ISM}}
\right)^{\frac{1}{2}} \left[ \frac{\gamma+3}{2(\gamma+1)}
\right]^{\frac{1}{2}}. \\
 \frac{R_{\rm HP}}{R'_{\rm TS}}&=1.58 \pm 0.05,
\label{RTS.incompr}
\end{align}
where $\rho_{obs}$ and $u_{obs}$ are respectively plasma density and plasma velocity measured \emph{in situ} at distance $R_{obs}$, $P_{\rm ISM}$ is the stagnation pressure discussed in section 4 of~\citet{2019HelMod}, and $\gamma=5/3$. $R'_{TS}$ is defined as the TS position at the time when it was left by the SW stream that is currently reaching the HP \citep[for more details see][]{2019HelMod}; this typically takes about 4 years, but depends on the SW speed. Therefore, the actual dimensions of the heliosphere used in \helmod{}-4 evolve with time. The predicted TS distances are in good agreement with those observed: for Voyager 1 (Voyager 2) the detected TS position is 93.8 AU (83.6 AU) and the predicted is 91.8 AU (86.3 AU), i.e.\ within 3 AU ($<$3.5\% error). Regarding the HP, based on the $R_{\rm HP}$ observed by Voyager 1, the predicted $R_{\rm HP}$ at the time of the Voyager 2 crossing was 120.7 AU while in reality it is 119 AU.  

In the present code, particular attention is paid to the quality of description of the high solar activity periods, which is evaluated though a comparison of \helmod{} calculations and the CR proton data by AMS-02 \citep{2018PhRvL.pHe}, and to transitions from/to solar minima. This was achieved through introduction of a drift suppression factor and particle diffusion parameters which depend on the level of solar disturbances~\citep[see a discussion in][]{2019HelMod}.

\subsection{Scenarios of a spectral break at 300 GV}\label{scenarios}

In 2011 PAMELA collaboration reported observation of a new break (hardening) in the spectra of the most abundant CR species, protons and He, above a rigidity of a few hundred GV \citep{2011Sci...332...69A}. This publication confirmed the hardening of the CR proton and He spectra found in earlier experiments, ATIC-2 \citep{2008ICRC....2...31W,2009BRASP..73..564P} and CREAM \citep{2010ApJ...714L..89A,2011ApJ...728..122Y}. Later, the break was also confirmed by the \fermilat{} \citep{2014PhRvL.112o1103A} and with much higher precision and statistics by AMS-02 \citep{2015PhRvL.114q1103A,2015PhRvL.115u1101A}. The break is smooth and observed at the same rigidity $\sim$300 GV for both species.

The interpretations of the break started to appear soon after the PAMELA publication. Perhaps the first was a paper by \citet{2012ApJ...752...68V}, which offered three distinctly different scenarios that can be tested through precise measurements of secondary species, such as secondary nuclei and $\bar{p}$, and anisotropy measurements. The proposed interpretations include (i) the ``propagation'' ({\it P}) scenario, where the observed break is the result of a change in the spectrum of interstellar turbulence that translates into a break in the index of the diffusion coefficient, (ii) the ``injection'' ({\it I}) scenario, where the break is due to the presence of populations of CR sources injecting particles with softer and harder spectra, and (iii) the ``local source'' scenario, where the local source injects low- ({\it L}-) or high-energy ({\it H}-scenario) particles with the spectrum that is correspondingly softer or harder than the rest of CRs produced by distant sources.

The {\it P}-scenario implies that the break should be observed in spectra of all CR species at about the same rigidity since the interstellar turbulence acts on all particles. Thus the index of the rigidity dependence of the diffusion coefficient $\delta$ has to have a break at the same rigidity, where its values $\delta_1$ below and $\delta_2$ above the break are connected with the observed value of the break $\Delta\alpha$ in the spectral index of \emph{propagated} primary species: $\delta_2=\delta_1-\Delta\alpha$, where $\Delta\alpha=\alpha_1-\alpha_2$, and $\alpha_1$ ($\alpha_2$) is the observed spectral index of primaries below (above) the break. The observed spectrum of secondary species has an index $\epsilon_1= \alpha_1+\delta_1$ below the break and an index $\epsilon_2= \alpha_2+\delta_2=\alpha_1+\delta_1-2\Delta\alpha$ above the break. Therefore, the change in the spectral index of secondary species would be $\Delta\epsilon=\epsilon_1 -\epsilon_2=2\Delta\alpha$, i.e., twice the value of the break observed in the spectra of primary species. This scenario also predicts an almost flat ratio $\bar{p}/p$ in the reacceleration model. Besides, the index $\delta_2$ of the rigidity dependence of the diffusion coefficient above the break becomes smaller and thus more consistent with CR anisotropy measurements \cite[see a collection of data in][]{2006AdSpR..37.1909P}.  

The {\it I}-scenario implies that the index of the diffusion coefficient has no break. Therefore, the change in the spectral index of secondary species would be the same as for primaries: $\Delta\epsilon=\Delta\alpha$. The break in the primary and secondary nuclei cancels when we take their ratio, as, e.g., the B/C ratio that would be exactly the same as if there is no break. In this scenario, the position of the break in the spectra of individual CR species may vary dependently on the composition of CR sources injecting particles at low and high energies, while the predicted anisotropy would exceed the actual measurements, see also \citet{2012PhPl...19h2901M}.

The local source scenario implies that the local source dominates in some part of the observed spectrum, at low ({\it L}-scenario) or high energies ({\it H}-scenario). Therefore, the amount of secondaries should drop significantly in the corresponding energy range since the freshly accelerated particles did not have time for fragmentation. A significant contribution of the local source at high energies ({\it H}-scenario) would also dramatically increase the CR anisotropy that may be in conflict with observations \cite[][]{2006AdSpR..37.1909P,2013APh....50...33S}.

\citet{2012ApJ...752...68V} conclude that the {\it P}-scenario is preferred, but the absence of reliable measurements of secondary species above a few hundred GV at that time did not allow to distinguish between different options.

Interestingly, soon after that publication, \citet{2012PhRvL.109f1101B} proposed a mechanism of the formation of the spectrum of interstellar turbulence. According to this model, the position of the break in the index of the diffusion coefficient correspond to the case when the diffusive propagation is no longer determined by the self-generated turbulence, but rather by the cascading of externally generated turbulence (for instance due to supernova bubbles) from large spatial scales to smaller scales. Independently on the origin of the break in the spectrum of interstellar turbulence, this would also lead to the expression $\Delta\epsilon=2\Delta\alpha$. There are also more recent papers that apply various versions of the {\it P}-scenario to the available data \citep[e.g.,][]{2017PhRvL.119x1101G,2019ApJ...873...77N}.

A subsequent release of the spectra of other primary and secondary species by AMS-02 was eagerly awaited. First, AMS-02 confirmed a clear distinction in the rigidity dependencies between the groups of mostly primary nuclei, He, C, O \citep{2017PhRvL.119y1101A}, and secondary nuclei, Li, Be, B \citep{2018PhRvL.120b1101A}, while the nuclei within each group have similar spectra. Nitrogen that is half-primary/half-secondary fell in between \citep{2018PhRvL.121e1103A}. The spectral index of the primary species, C, O, below/above the break is about $\alpha_1/\alpha_2$$\approx$2.65/2.55, while for secondary species it is $\epsilon_1/\epsilon_2$$\approx$3.1/2.9. Here the indices below/above the break were taken at $\approx$50 GV/700 GV, where the solar modulation is negligible. One can see that the change in the spectral index of secondaries $\Delta\epsilon$$\approx$0.2 is double the change in the spectral index of primaries $\Delta\alpha$$\approx$0.1, as predicted in the {\it P}-scenario \citep{2012ApJ...752...68V} long before the data on secondary species became available \citep{2018PhRvL.120b1101A}. Note that because there are many different types of sources of CR electrons and positrons and due to the large energy losses of these particles, the spectra of electrons and positrons may behave quite differently and they indeed do so \citep{2019PhRvL.122d1102A,2019PhRvL.122j1101A}. 

\section{Numerical Procedure}

\begin{deluxetable*}{rcccc}
	\tablewidth{0mm}
	\tablecaption{Injection spectra of primary species \label{tbl-inject}}
	\tablehead{
		\colhead{}& \multicolumn{3}{c}{Spectral parameters}\\ 
		\colhead{Isotope}&
		\colhead{$\gamma_0\>\> ^{\displaystyle R_0 {\rm (GV)}} \> s_0$} &
		\colhead{$\gamma_1\>\> ^{\displaystyle R_1 {\rm (GV)}} \> s_1$} & 
		\colhead{$\gamma_2\>\> ^{\displaystyle R_2 {\rm (GV)}} \> s_2$} & 
		\colhead{$\gamma_3$}}
	\startdata
	$^{1,2}$H\phm{e} & 
	$2.35\>\> ^{\displaystyle 1.15}\> 0.20$ & 
	$1.71\>\> ^{\displaystyle 6.90}\> 0.23$ & 
	$2.44\>\> ^{\displaystyle 365}\> 0.09$ &
	2.25 \\
	
	$^{3,4}$He & 
	$2.24\>\> ^{\displaystyle 1.00}\> 0.20$ & 
	$1.83\>\> ^{\displaystyle 7.30}\> 0.22$ & 
	$2.40\>\> ^{\displaystyle 325}\> 0.16$ &
	2.15 \\
	
	$^{7}$Li\tablenotemark{a} & 
	\nodata & 
	$1.10\>\> ^{\displaystyle 12.0}\> 0.16$ & 
	$2.72\>\> ^{\displaystyle 355}\> 0.13$ &
	1.90 \\
	
	$^{12,13}$C\phm{e} & 
	$1.00\>\> ^{\displaystyle 0.95}\> 0.16$ & 
	$2.01\>\> ^{\displaystyle 6.10}\> 0.29$ & 
	$2.42\>\> ^{\displaystyle 340}\> 0.15$ &
	2.12 \\
	
	$^{14}$N\phm{e} & 
	$1.13\>\> ^{\displaystyle 1.20}\> 0.15$ & 
	$1.98\>\> ^{\displaystyle 7.00}\> 0.20$ & 
	$2.44\>\> ^{\displaystyle 300}\> 0.15$ &
	1.87 \\
	
	$^{16,18}$O\phm{e} & 
	$1.11\>\> ^{\displaystyle 1.20}\> 0.19$ & 
	$1.99\>\> ^{\displaystyle 7.70}\> 0.33$ & 
	$2.46\>\> ^{\displaystyle 365}\> 0.15$ &
	2.13 \\
	
	Others &
	$1.12\>\> ^{\displaystyle 1.10}\> 0.16$ & 
	$1.97\>\> ^{\displaystyle 7.00}\> 0.19$ & 
	$2.44\>\> ^{\displaystyle 355}\> 0.15$ &
	2.13
	\enddata
	\tablenotetext{a}{Models with primary lithium.}
	\tablecomments{For definitions of the injection parameters see Eq.~(\ref{eq:1}). The fit errors: $\gamma_{0,1} \pm0.06$, $\gamma_{2,3} \pm0.04$, $R_0 \pm0.5$~GV, $R_1\pm1$ GV, and $R_2\pm15$ GV. In the case of the {\it P}-scenario, the parameter $\gamma_3$ is not used (see Section~\ref{scenarios} for details).}
\end{deluxetable*}

\begin{figure*}[tbh!]
	\centering
	\includegraphics[width=0.98\textwidth]{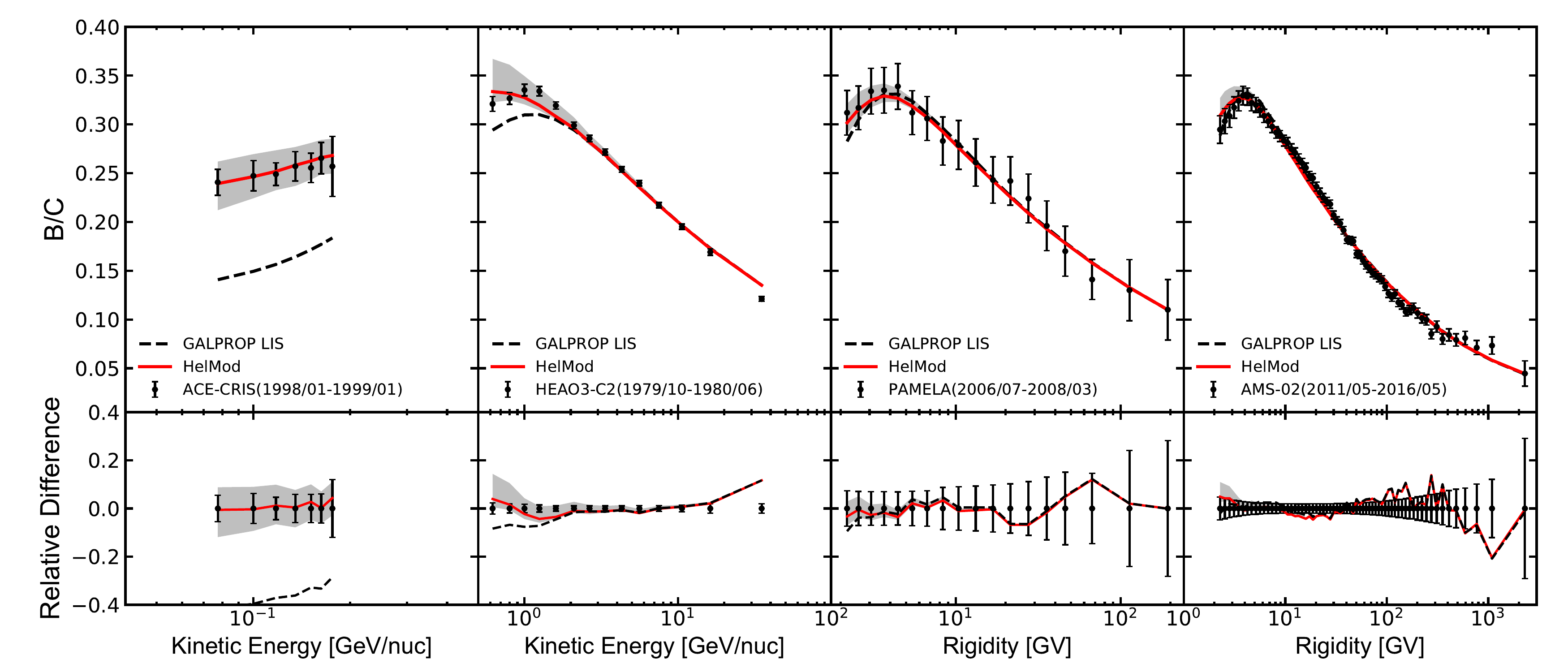}
	\caption{Top panels show the B/C ratio as measured by (from left to right) ACE/CRIS \citep{2013ApJ...770..117L}, HEAO-3 \citep{1990A&A...233...96E}, PAMELA \citep{2014ApJ...791...93A}, and AMS-02 \citep{PhysRevLett.120-2018} compared to our calculations. The dashed black line shows the \galprop{} LIS ratio ({\it I}-scenario), the modulated ratios (that correspond to each individual period of data taking) are shown by the red lines. Bottom panels show the relative difference between the calculations and a corresponding experimental data set.}
	\label{fig:BoC}
\end{figure*}

\begin{deluxetable}{lrlcc}
	\tablewidth{0mm}
	\tablecaption{Best-fit propagation parameters for {\it I}- and {\it P}-scenarios\label{tbl-prop}}
	\tablehead{
		\colhead{N}& \colhead{Parameter}& \colhead{Units}& \colhead{Best Value}& \colhead{Error} 
	}
	\startdata
	\phn1 & $z_h$ & kpc &4.0 &0.6\\
	\phn2 & $D_0 (R= 4\ {\rm GV})$ & $10^{28}$ cm$^{2}$ s$^{-1}$  &4.3 &0.7\\
	\phn3\tablenotemark{a} & $\delta_1$ & &0.415 &0.025\\
	\phn4 & $V_{\rm Alf}$ & km s$^{-1}$ &30 &3\\
	\phn5 & $dV_{\rm conv}/dz$ & km s$^{-1}$ kpc$^{-1}$ & 9.8 &0.8
	\enddata
	\tablenotetext{a}{For the {\it P}-scenario (see Section \ref{scenarios}): $\delta_2=0.15\pm 0.03$ for $R\ge370\pm 25$ GV.}
\end{deluxetable}

To derive the LIS of CR species we use the same optimization procedure that was employed in our previous analyses \citep{2017ApJ...840..115B,2018ApJ...854...94B,2018ApJ...858...61B}. The combined framework, described in \citet{2017ApJ...840..115B}, is logically divided into two parts:  (i) a MCMC interface to version 56 of \galprop{}~\citep{MasiDM2016}, that allows for sampling of the production and propagation parameters space, and (ii) an iterative procedure that, starting from \galprop{} output, provides modulated spectra computed with \helmod{} to compare with AMS-02 data as observational constraints \citep{2018AdSpR..62.2859B}. The final product is a set of Galactic and heliospheric propagation parameters for all CR species to determine the LIS that best reproduces the available experimental data.

The basic properties of CR propagation in the Galaxy are described by the transport equations quite well, but the exact values of the propagation parameters depend on the assumed propagation model and accuracy of selected CR datasets. Therefore, we used the MCMC procedure to determine the propagation parameters using the best available CR data. Five main propagation parameters, that affect the overall shape of CR spectra, were left free in the scan using \galprop{} running in the 2D mode: the Galactic halo half-width $z_h$, the normalization of the diffusion coefficient $D_0$ at the reference rigidity $R=4$ GV and the index of its rigidity dependence $\delta$, the Alfv\'en velocity $V_{\rm Alf}$, and the gradient of the convection velocity $dV_{\rm conv}/dz$ ($V_{\rm conv}=0$ in the plane, $z=0$). The radial size of the Galaxy does not significantly affect the values of propagation parameters and was set to 20 kpc. Besides, we introduced a factor $\beta^\eta$ in the diffusion coefficient, where $\beta=v/c$, and $\eta$ was left free. The best fit value of $\eta=0.71$ improves the agreement at low energies, and slightly affects the choice of injection indices ${\gamma}_0$ an ${\gamma}_1$ (Table~\ref{tbl-inject}).

It is worth mentioning that simultaneous inclusion of both distributed reacceleration and convection is necessary to describe the high precision AMS-02 data, particularly in the range below 10 GV where they significantly affect the spectra of CR species \citep[for more details see][]{2017ApJ...840..115B}. The best-fit values of the main propagation parameters tuned to AMS-02 data are listed in Table~\ref{tbl-prop}, which are about the same as obtained in \citet{2017ApJ...840..115B}, within the quoted error bars. The most significant change is a slight increase of the Alfv\'en velocity $V_{\rm Alf}$ that improves an agreement with the B/C ratio and electron data \citep{2018ApJ...854...94B}.

The MCMC procedure is used only in the first step to define a consistent parameter space, then a methodical calibration of the model employing the \helmod{} module was performed. Parameters of the injection spectra, such as spectral indices $\gamma_i$ and the break rigidities $R_i$, were also left free, but their exact values depend on the solar modulation, so the low energy parts of the spectra are tuned together with the solar modulation parameters. The modulated spectra of CR protons are used as a reference for evaluation of the modulation parameters assuming that all Galactic CRs species are subject to the same heliospheric conditions in the considered energy range. The best fit injection parameters are listed in Table \ref{tbl-inject}, see also Eq.~(\ref{eq:1}) for definitions.

\begin{figure}[tb!]
	\centering
	\includegraphics[width=0.49\textwidth]{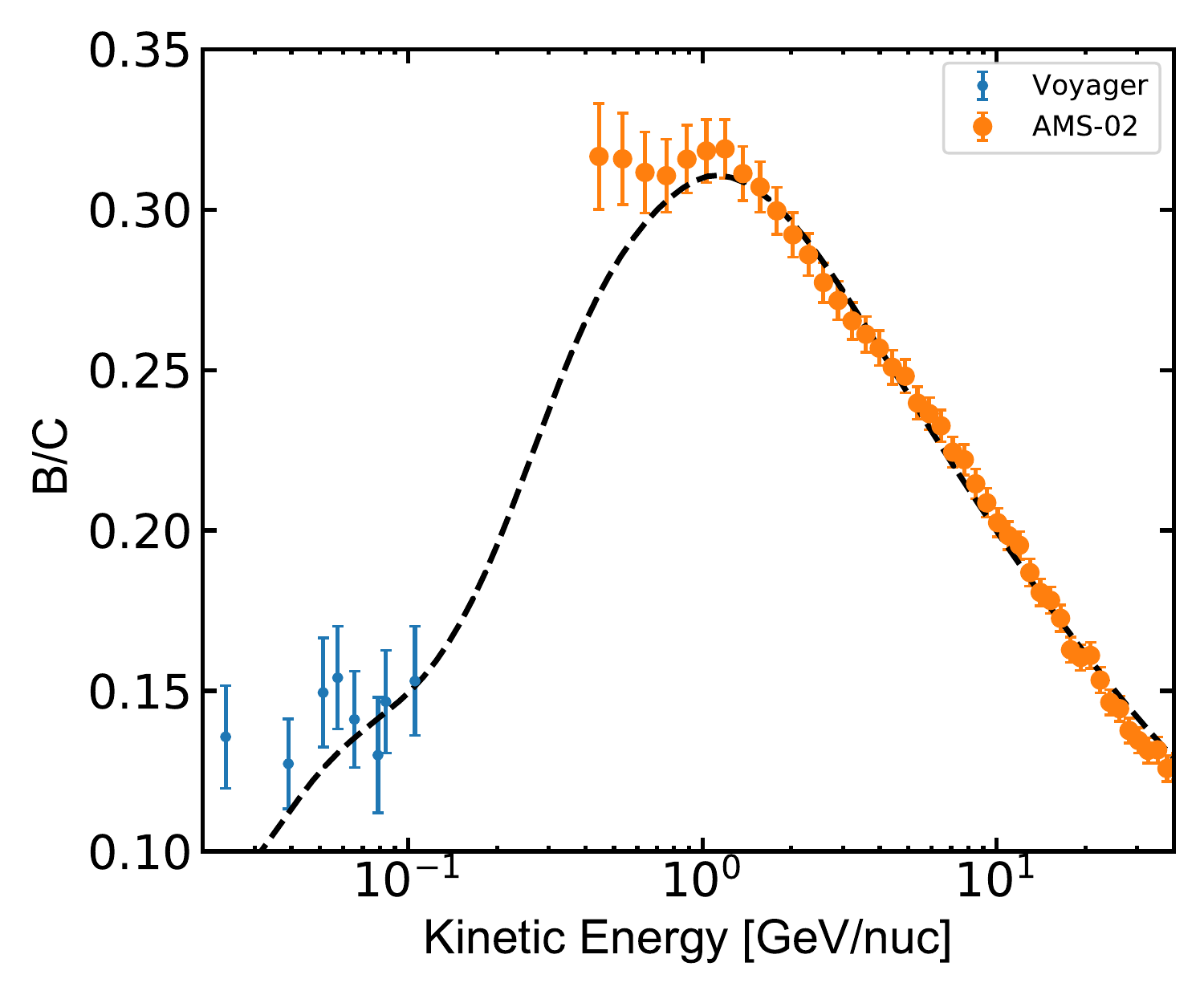}
	\caption{The B/C ratio as measured by  AMS-02 \citep{PhysRevLett.120-2018} and by Voyager 1  \citep{2016ApJ...831...18C} reported in kinetic energy per nucleon.  The dashed black line shows the \galprop{} LIS ratio. AMS-02 data in kinetic energy per nucleon are those obtained from ASI Cosmic Ray Database \citep{2017ICRC...35.1073D}.}
	\label{fig:BoC_Voy}
\end{figure}

Our calculations of the B/C ratio in the diffusion-convection model are shown in Figures~\ref{fig:BoC}, \ref{fig:BoC_Voy} along with the data by AMS-02 \citep{PhysRevLett.120-2018}, PAMELA \citep{2014ApJ...791...93A}, HEAO-3 \citep{1990A&A...233...96E}, ACE/CRIS \citep{2013ApJ...770..117L}, and Voyager \citep{2016ApJ...831...18C}. The model and calculations are described in detail in \citet{2017ApJ...840..115B}. The agreement is good for all instruments and all epochs given only one LIS set is inferred from the AMS-02 data.

\section{Results}\label{results}

The results of our calculations are compared with available data from AMS-02 \citep{2018PhRvL.120b1101A,2018PhRvL.121e1103A}, ACE/CRIS \citep{2006AdSpR..38.1558D}, PAMELA \citep{2014ApJ...791...93A}, HEAO-3 \citep{1990A&A...233...96E}, and Voyager 1 \citep{2016ApJ...831...18C}.  The LIS for pure secondary species, such as beryllium and boron, were defined assuming LIS spectra for carbon and oxygen as derived in \citet{2018ApJ...858...61B} and discussed in the next section. Lithium LIS was initially evaluated using the same procedure. This leads to a large deviation, about 20\% excess in the broad range of rigidities from 5 GV -- 1 TV, when compared to the AMS-02 data: this anomaly is discussed in detail in Section \ref{Lithium}. In this work we also consider LIS for nitrogen that is approximately half-primary/half-secondary.

\subsection{Beryllium, Boron, and Nitrogen} \label{Sect:Results_2}

\begin{figure}[tb!]
	\centering
	\includegraphics[width=0.49\textwidth]{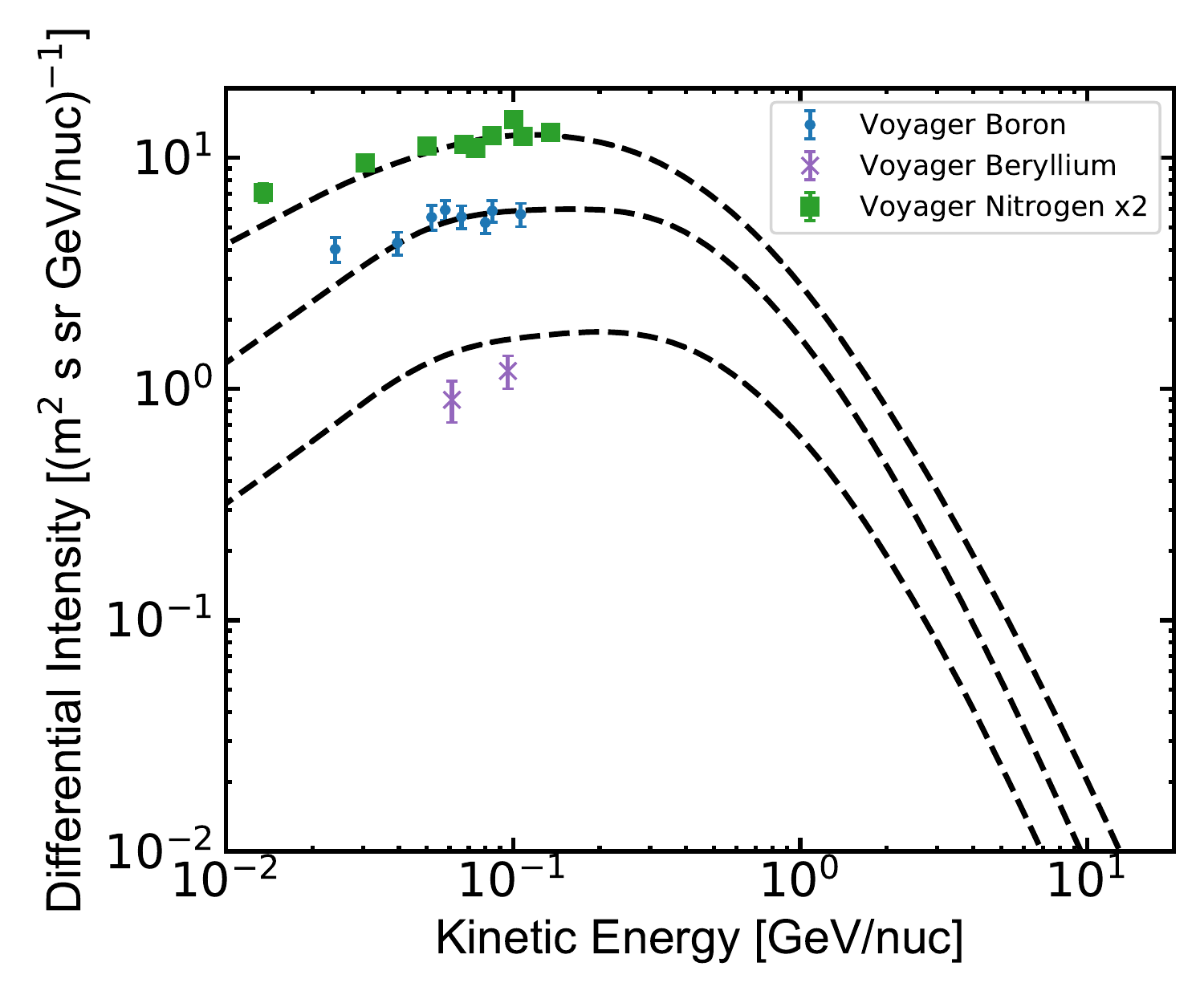}
	\caption{The \galprop{} LIS spectra (black dashed lines) are compared to the Voyager 1 data \citep{2016ApJ...831...18C}: boron -- small blue dots, beryllium -- violet crosses, and nitrogen -- green squares. Nitrogen spectrum is scaled up by a factor of 2 for clarity.}
	\label{fig:BeBN_LIS}
\end{figure}

\begin{figure*}[tb!]
	\centering
	\includegraphics[width=0.98\textwidth]{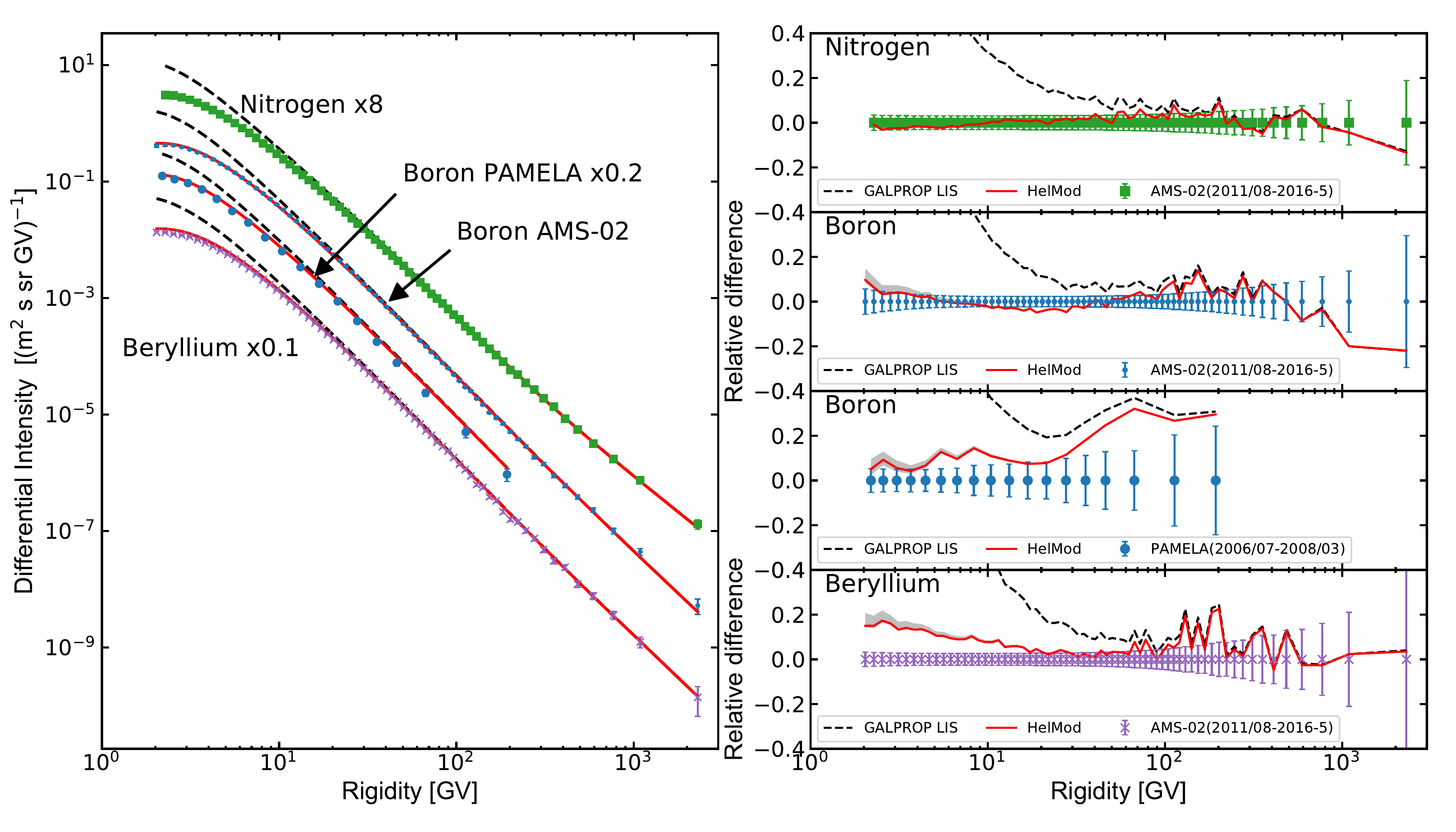}
	\caption{Our model calculations are compared to the observed CR spectra of nitrogen (green squares), boron (blue dots), and beryllium (violet crosses) from AMS-02 \citep{2018PhRvL.120b1101A,2018PhRvL.121e1103A} and PAMELA \citep[boron -- filled blue circles,][]{2014ApJ...791...93A}. Black dashed line shows the LIS spectra ({\it I}-scenario), the red solid lines are modulated to the level that corresponds to the period of data taking. For clarity, the AMS-02 spectra of nitrogen and beryllium, and PAMELA spectrum of boron are scaled by factors 8, 0.1, and 0.2, respectively. The right panel show the relative difference between calculations and a corresponding experimental data set.}
	\label{fig:BeBN_1}
\end{figure*}

\begin{figure*}[tb!]
	\centering
	\includegraphics[width=0.98\textwidth]{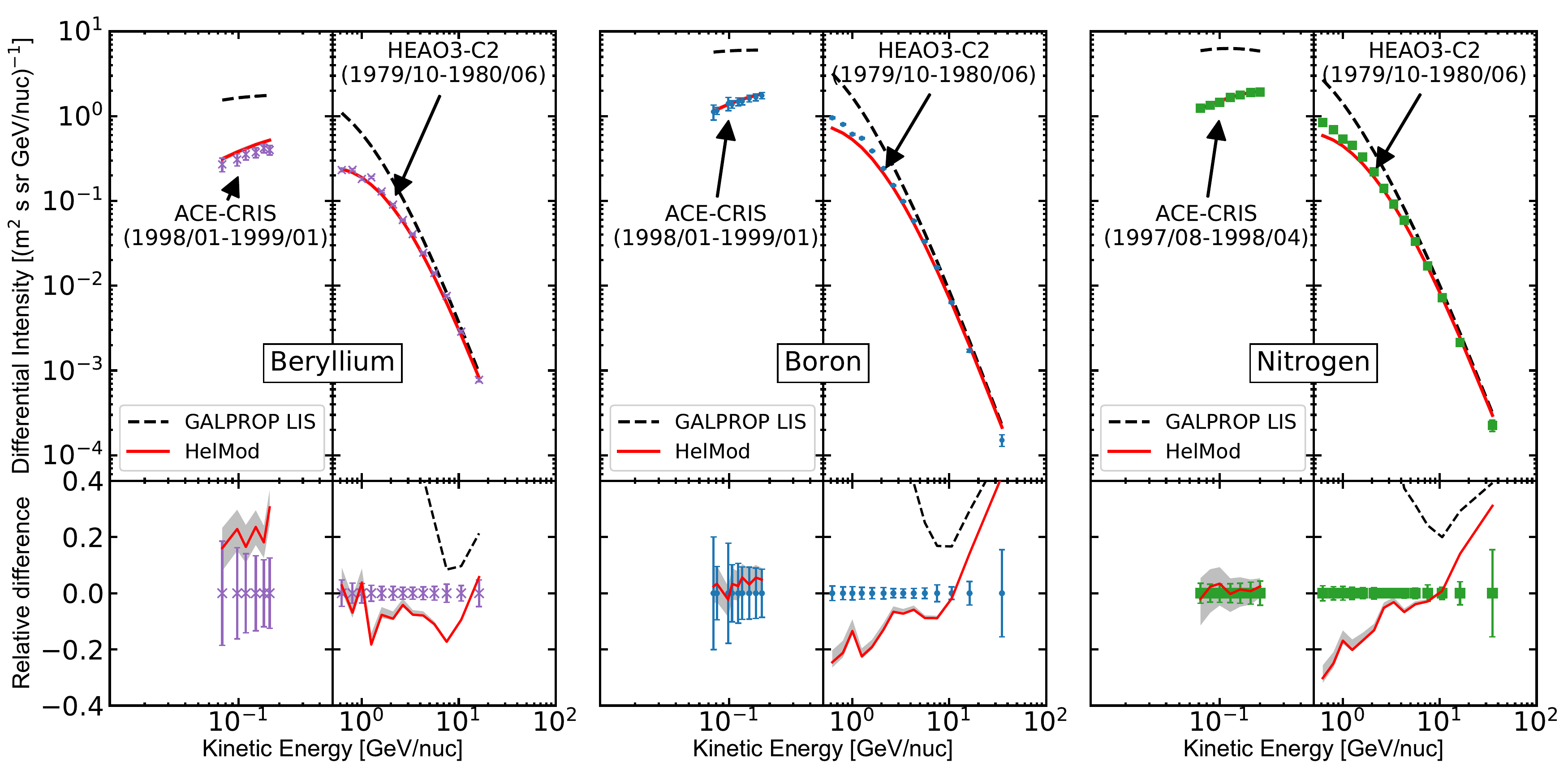}
	\caption{Our model calculations are compared to the observed CR spectra of beryllium (violet crosses, left panel), boron (blue points, center panel), and nitrogen (green squares, right panel) from ACE-CRIS \citep{2006AdSpR..38.1558D} and HEAO3-C2 \citep{1990A&A...233...96E}. Black dashed lines show the LIS spectra, the red solid lines are modulated to the levels that correspond to the periods of data taking. Bottom panels show the relative difference between the calculations and a corresponding experimental data set.}
	\label{fig:BeBN_2}
\end{figure*}

The results of our calculations in the {\it I}-scenario, the LIS of beryllium, boron, and nitrogen, are shown in Figures \ref{fig:BeBN_LIS}-\ref{fig:BeBN_2} as compared to the available data. A comparison of our calculations with data taken at different levels of solar activity and at different polarities of the solar magnetic field demonstrates overall good agreement thus supporting our model approach. 

A moderate overprediction of 10\%--20\% (at 2--10 GV) is observed in the case of beryllium (Figure \ref{fig:BeBN_1}) when compared to AMS-02 data taken during the same period. Some overprediction ($\sim$20\%) is also observed when comparing to the Voyager 1 (Figure \ref{fig:BeBN_LIS}) and ACE/CRIS data (Figure \ref{fig:BeBN_2}), but in the latter cases it is at the level of $\la$$2\sigma$. The overprediction in beryllium flux below a few GV is most likely connected with errors in the total inelastic cross sections of beryllium isotopes. The cross section errors are most significant in the case of beryllium and, in particular, in the energy range below 10 GeV/n \citep[see Figure 6 in][]{2018PhRvC..98c4611G}.    

Other discrepancies, such as those reported in Figure \ref{fig:BeBN_2} for beryllium, boron, and nitrogen when comparing to the HEAO-3 data \citep{1990A&A...233...96E}, are an indication of the systematic uncertainties of the instrument. This is demonstrated by the excellent agreement of spectra of boron and nitrogen with AMS-02 data \citep{2018PhRvL.120b1101A} in the whole energy range from 2 GV -- 2 TV.

\begin{figure}[tb!]
	\centering
	\includegraphics[width=0.49\textwidth]{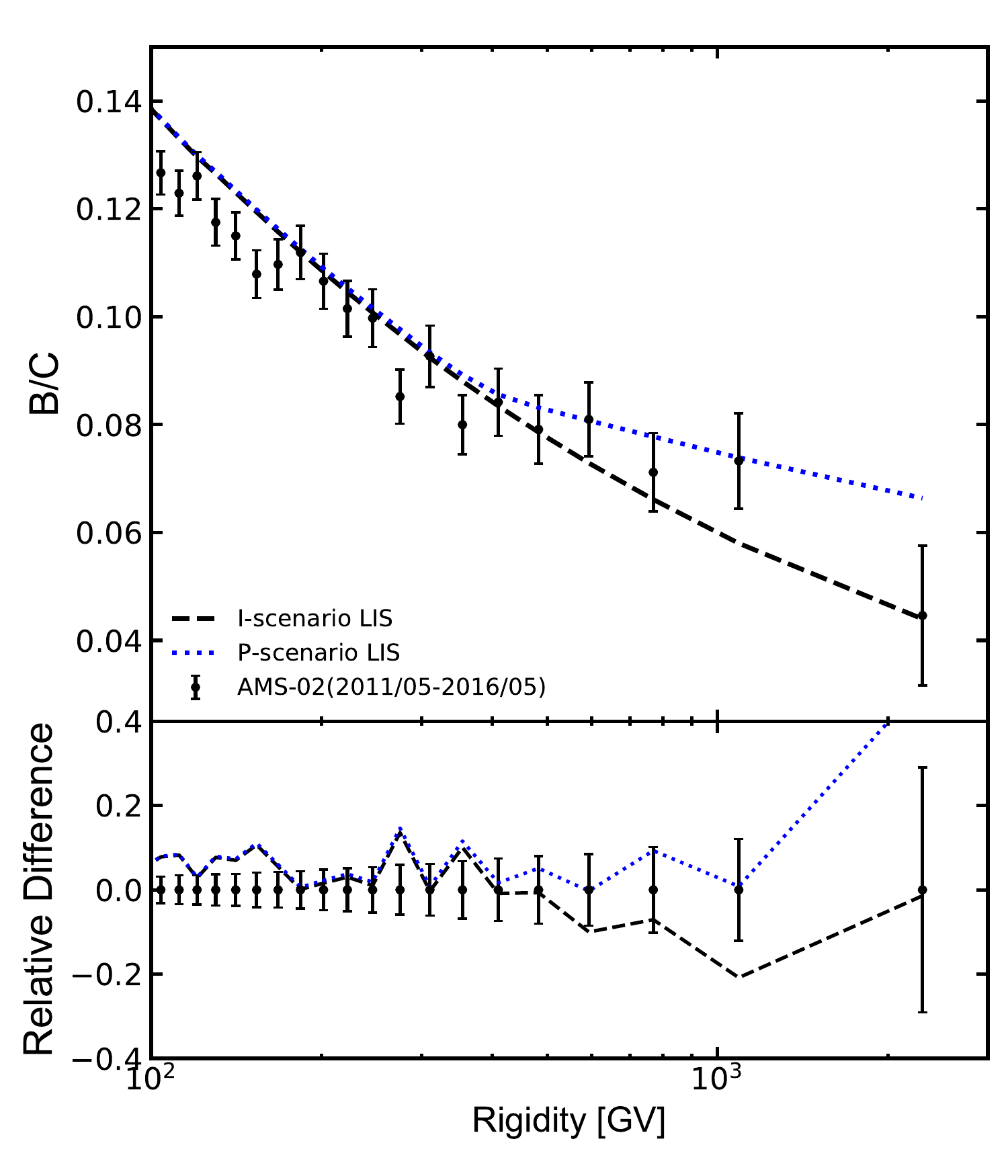}
	\caption{Our model calculations of the B/C ratio in the $I$-scenario (dashed black line) and $P$-scenario (blue dotted line) are compared to the AMS-02 data \citep{PhysRevLett.120-2018}. Bottom panel shows the relative difference between the calculations and a corresponding experimental data set.}
	\label{fig:BoC_Break}
\end{figure}

\begin{figure}[tb!]
	\centering
	\includegraphics[width=0.49\textwidth]{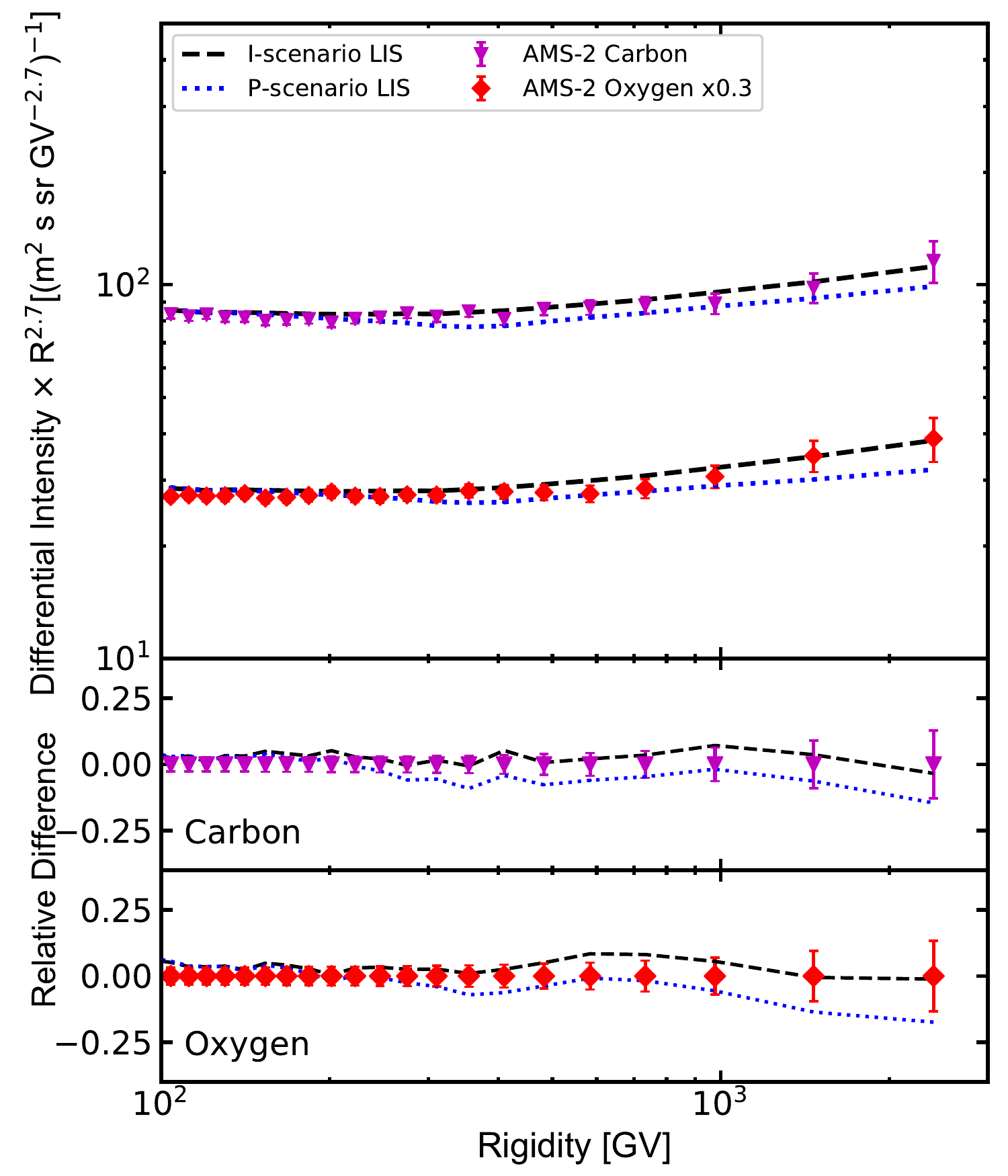}
	\caption{Our model calculations of the carbon and oxygen spectra in the $I$-scenario (dashed black line) and $P$-scenario (blue dotted line) are compared to the AMS-02 data \citep{2017PhRvL.119y1101A}. The experimental data points for carbon are shown by violet triangles, and for oxygen by red diamonds. The differential intensity is multiplied by $R^{2.7}$ to enhance the high energy differences. Oxygen spectrum is scaled by a factor 0.3 for clarity. Bottom panels show the relative difference between the calculations and a corresponding experimental data set.}
	\label{fig:AMS02_Break_2}
\end{figure}

Our results for the {\it I}- and {\it P}-scenarios are shown in Figures \ref{fig:BoC_Break}-\ref{fig:AMS02_Break_1}. Since the spectral behavior of CR nuclei at low energies is the same in both scenarios \citep[also in plots in our earlier papers][]{2017ApJ...840..115B,2018ApJ...858...61B}, the {\it P}-scenario calculations are only shown above $\sim$100 GV where the effects of solar modulation can be safely neglected. Figure \ref{fig:BoC_Break} shows the B/C ratio in the {\it P}-scenario together with AMS-02 data taken at different epochs. The flattening above $\sim$350 GV is clearly seen, but it is very moderate and agrees well with data.

The spectra of all nuclei species, Be, B, C, N, O, calculated in both scenarios are shown in Figures \ref{fig:AMS02_Break_2}, \ref{fig:AMS02_Break_1}. In both cases the agreement with data is good. In the case of {\it P}-scenario we use only parameters $\gamma_{0,1,2}$ and $R_{0,1}$ from Table \ref{tbl-inject}, while the break at $\sim$350 GV appears due to the break in the diffusion coefficient (Table \ref{tbl-prop}). 

The effect of the local source producing low-energy CRs ({\it L}-scenario) on the observed composition of CRs and production of secondaries was studied in \citet{2003ApJ...586.1050M} and \citet{2012ApJ...752...68V}. In short, tuning to the observed secondary-to-primary ratio, such as B/C, would require a significant decrease in the normalization of the diffusion coefficient in order to compensate for the presence of locally produced primary species, such as C and O. This, in turn, would lead to an increase of the $\bar{p}/p$ ratio. However, a detailed exploration of this scenario is beyond the scope of this work. 

\subsection{Lithium Anomaly} \label{Lithium} \label{Li}

\begin{figure}[tb!]
	\centering
	\includegraphics[width=0.49\textwidth]{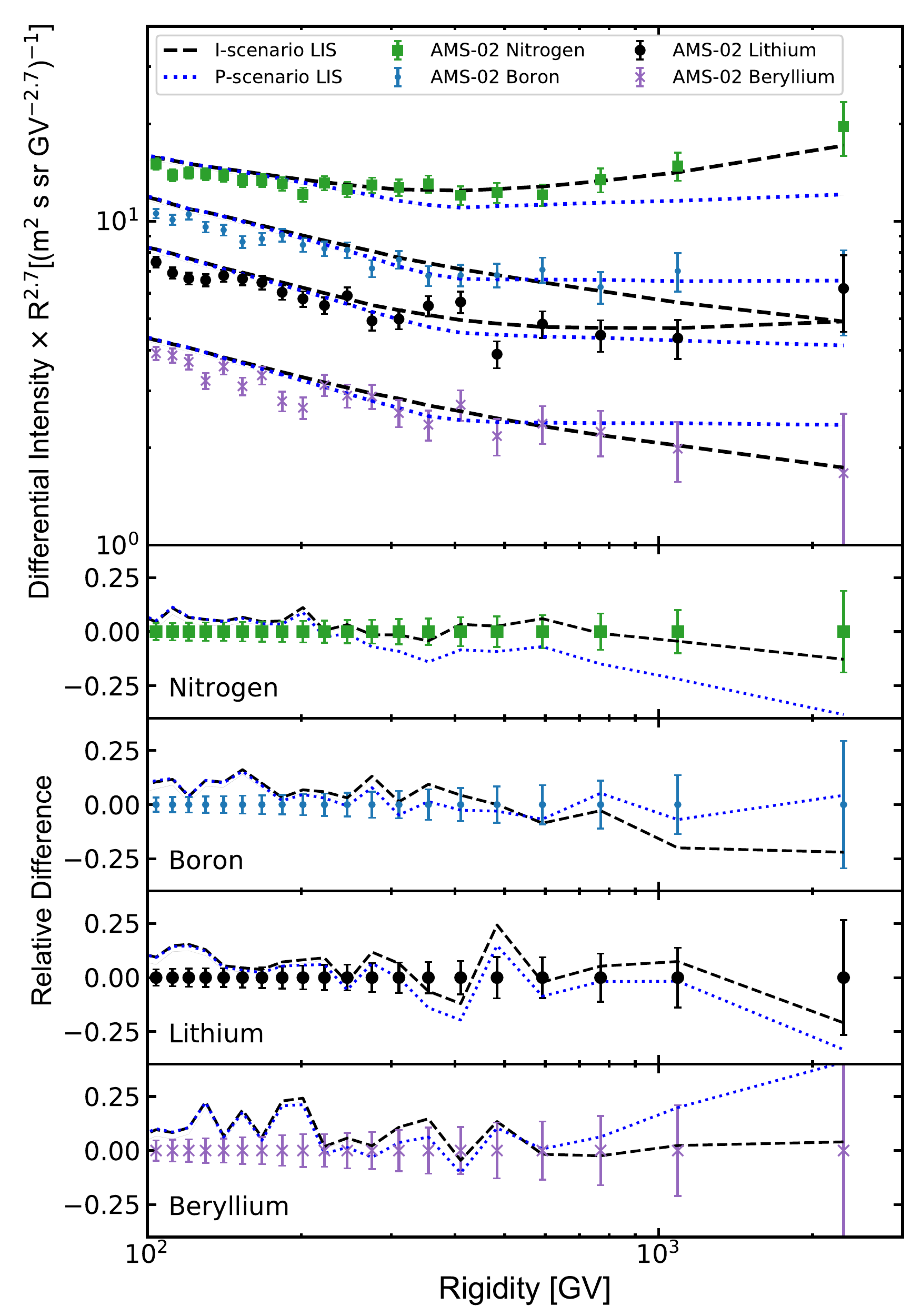}
	\caption{Our model calculations of the nitrogen, boron, lithium, and beryllium spectra in the $I$-scenario (dashed black line) and $P$-scenario (blue dotted line) as compared to the AMS-02 data \citep{PhysRevLett.120-2018}. The experimental data points are: nitrogen -- green squares, boron -- blue points, lithium -- black filled circles, and beryllium -- violet crosses. The differential intensity is multiplied by $R^{2.7}$ to enhance the high energy differences. Bottom panels show the relative difference between the calculations and a corresponding experimental data set.}
	\label{fig:AMS02_Break_1}
\end{figure}

\begin{figure}[tb!]
	\centering
	\includegraphics[width=0.49\textwidth]{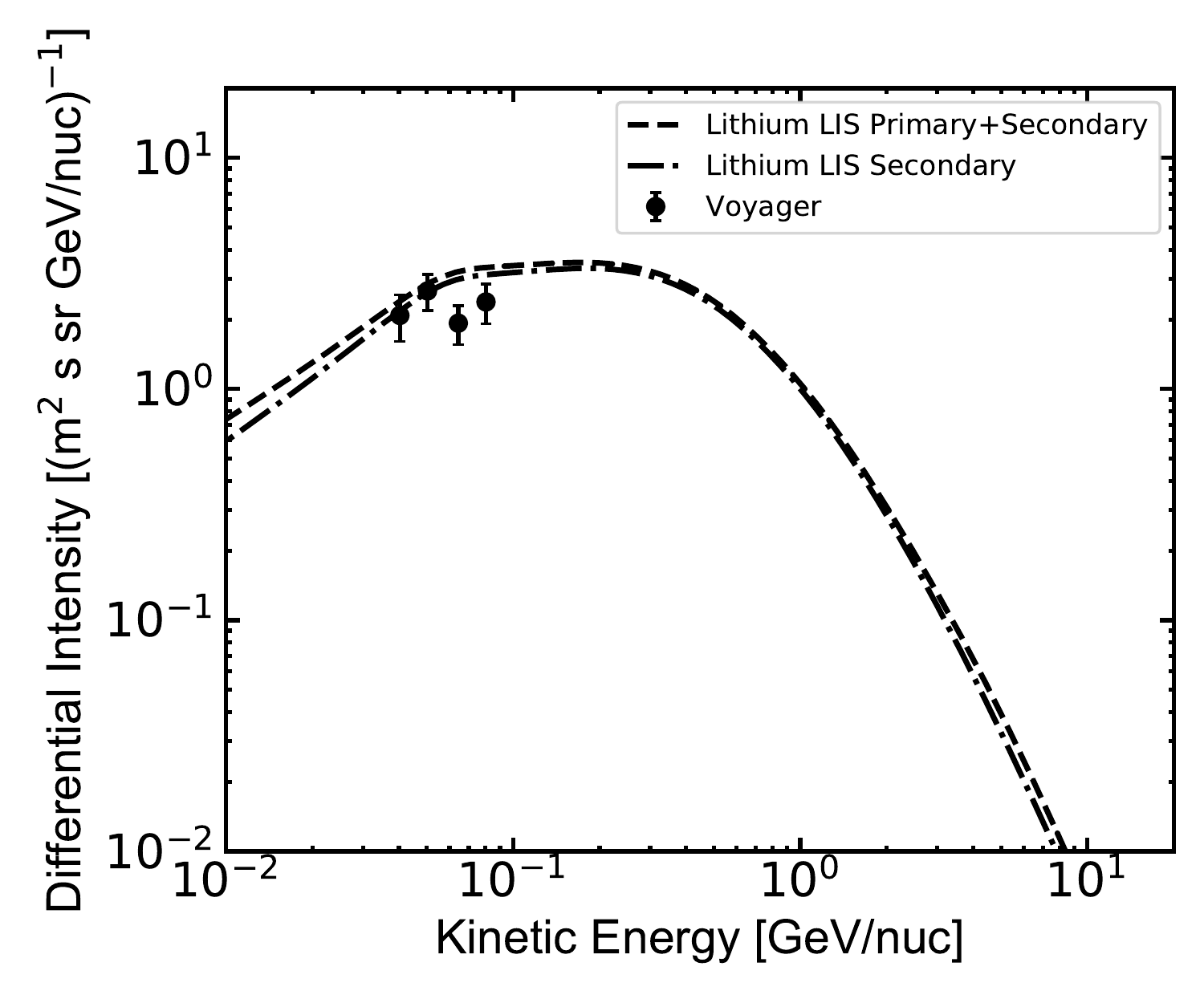}
	\caption{Our calculated spectra of secondary lithium (dot-dashed line) and lithium with added primary $^7$Li component (dashed line) are compared to the Voyager 1 data \citep{2016ApJ...831...18C}.}
	\label{fig:Li_LIS}
\end{figure}

\begin{figure}[tb!]
	\centering
	\includegraphics[width=0.49\textwidth]{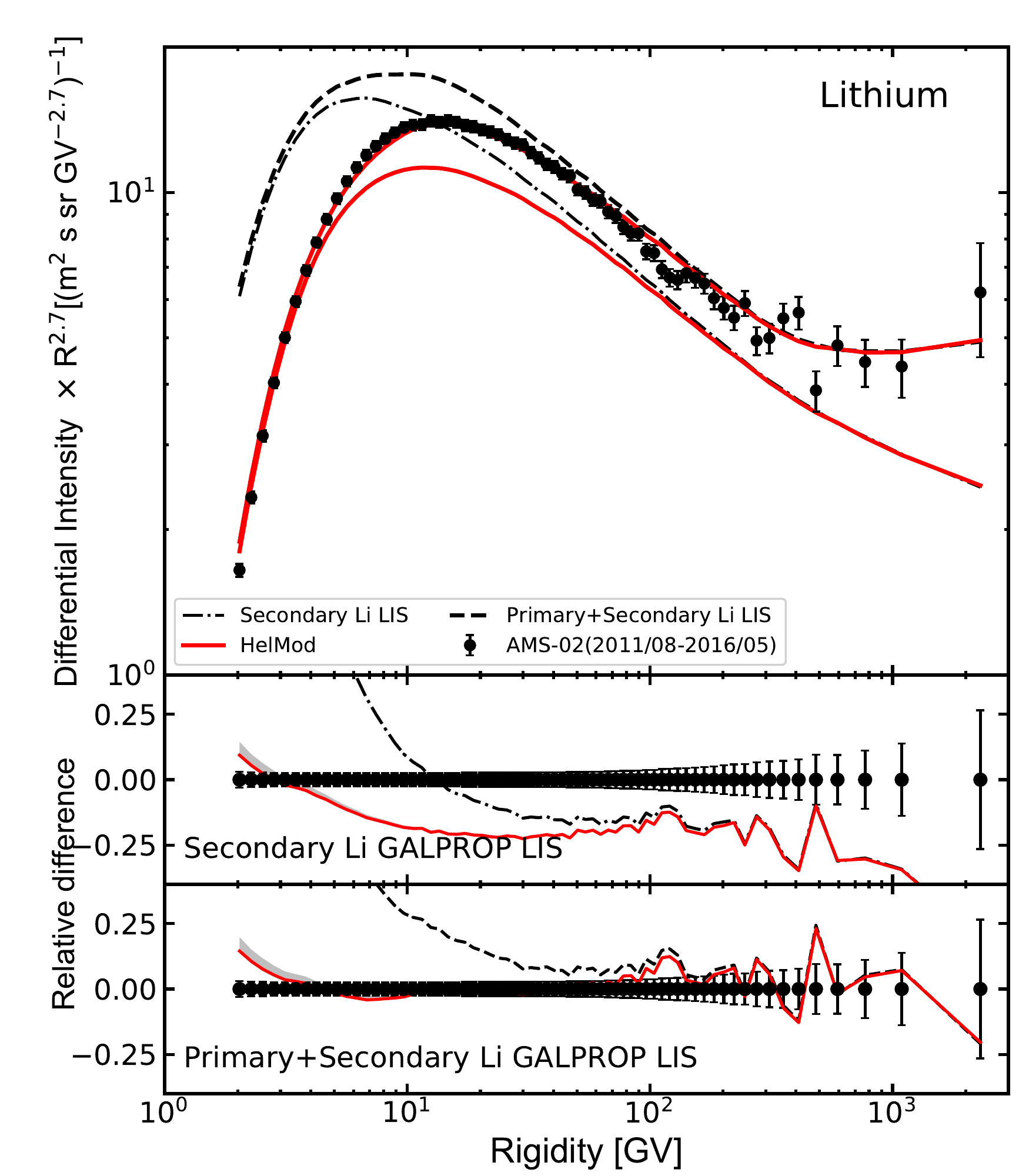}
	\caption{Our model calculations of CR lithium (in the $I$-scenario) as compared to the AMS-02 data \citep{2018PhRvL.120b1101A}. Black dot-dashed line shows only secondary component. Black dashed line shows the calculations that include the primary $^7$Li component. The red solid lines are modulated to the level that corresponds to the period of data taking. In the top panel, the differential intensity is multiplied by $R^{2.7}$ to enhance high energy differences.}
	\label{fig:Li_1}
\end{figure}

A demonstrated good agreement of our model calculations with measurements of CR species in a wide energy range implies that lithium spectrum should also be well reproduced by the same model. However, a comparison of our calculations of secondary lithium with data exhibits a significant excess over the model predictions above a few GV (Figures \ref{fig:Li_LIS}, \ref{fig:Li_1}) even though the propagation parameters are tuned to the B/C ratio. The additional \emph{secondary} $^7$Li cannot come from the decay of $^7$Be isotope in CRs that decays via K-capture. All \galprop{} calculations are already run with the processes of electron pickup, stripping, and K-capture included. Therefore, this may indicate that some part of the observed lithium has a different origin. 

Meanwhile, a possibly connected puzzle is a long-standing problem of the origin of lithium observed in today's universe  \citep[e.g.,][]{2015Natur.518..307H}. While some fraction of lithium was created in the primordial nucleosynthesis, most of the observed lithium is produced through interactions of energetic CR particles with interstellar gas (spallation reactions). On the other hand, the observed stellar lithium abundances indicate that some proportion of lithium is also produced in low-mass stars and nova explosions. Indeed, the alpha-capture reaction of $^7$Be production $^3$He$(\alpha,\gamma)^7$Be was proposed a while ago \citep{1955ApJ...121..144C,1971ApJ...164..111C}. A subsequent decay of $^7$Be with a half-life of 53.22 days yields $^7$Li isotope. To ensure that produced $^7$Li is not destroyed in subsequent nuclear reactions, $^7$Be should be transported into cooler layers where it can decay to $^7$Li, the so-called Cameron-Fowler mechanism. 

The production of $^7$Li in the same reactions in novae was first discussed by \citet{1975A&A....42...55A} and \citet{1978ApJ...222..600S}, while the details of the process were established later \citep{1996ApJ...465L..27H}. The amount of $^7$Li produced by a classical CO nova corresponds to about $10^{-10}M_\odot -10^{-9}M_\odot$ although the exact amount depends on many details of the explosion process. 

Recent observation of blue-shifted absorption lines of partly ionized $^7$Be in the spectrum of a classical nova V339 Del about 40-50 days after the explosion \citep{2015Natur.518..381T} is the first observational evidence that the mechanism proposed in 1970s is working indeed \citep{2015Natur.518..307H}. The observed blue shift of the absorption lines corresponds to the velocity of the ejecta reaching 1100--1270 km s$^{-1}$. Consequent observations of other novae (V1369 Cen -- \citealt{2015ApJ...808L..14I}, V5668 Sgr and V2944 Oph -- \citealt{2016ApJ...818..191T}, ASASSN-16kt [V407 Lupi] -- \citealt{2018MNRAS.478.1601I}, V838 Her -- \citealt{2018MNRAS.481.2261S}) also reveal the presence of $^7$Be lines in their spectra testifying that classical novae is the new type of sources of $^7$Li. The total mass of produced $^7$Li in these novae is estimated from $10^{-9}M_\odot-6\times 10^{-9}M_\odot$.

For our case this means that there is a source of primary lithium and that this lithium may be observed in CRs for the first time, thanks to the high statistics and precision achieved by the AMS-02 experiment \citep{2018PhRvL.120b1101A}. Though the absolute mass of lithium produced by novae is relatively small, the lithium abundance of the ejecta is comparable to that found in CRs given that the total mass of the ejecta is $10^{-5}M_\odot -10^{-4}M_\odot$. Such lithium-enriched ejecta could serve as a source of primary lithium that is subsequently accelerated in nova and SN shocks thus supplying additional lithium to CRs. 

We note that a similar mechanism was proposed by \citet{2018PhRvL.120d1103K} albeit with rather extreme assumption that the spectral hardening observed above 300 GV is due to the local SN Ia. In this model most of the protons, He, and lithium above the break are coming from the local source. This is equivalent to the {\it H}-scenario originally considered in \citet{2012ApJ...752...68V} and deemed unlikely, since it would lead to a dramatic increase in CR anisotropy at very high energies contrary to the observations, and a sharp drop in the B/C ratio above the break that is not observed either (see also a discussion in Section \ref{scenarios}).

Figure \ref{fig:Li_1} shows a fit ({\it I}-scenario) where a small amount of primary $^7$Li was injected with parameters given in Table~\ref{tbl-inject}, in addition to the secondary lithium. The agreement with the AMS-02 data \citep{2018PhRvL.120b1101A} is significantly improved. A small discrepancy around 2 GV is still remaining that may indicate some unknown systematics. A calculation in the {\it P}-scenario looks similarly and is not shown in this plot.

Our calculations of the lithium spectrum in the {\it P}-scenario shown in Figure \ref{fig:AMS02_Break_1} already contain primary $^7$Li that improves the agreement significantly. In this scenario, the injection spectrum of primary $^7$Li is the same as in the case of {\it I}-scenario (Table~\ref{tbl-inject}), but the parameter $\gamma_3$ in not used. This calculation testifies that the {\it P}-scenario is consistent with all data available from AMS-02 and earlier experiments and seems like a natural explanation of the break in the spectra of all CR nuclei species observed at $\sim$350 GV. Note that the {\it I}-scenario provides a similar quality fit, but at the cost of additional parameters. 

Another possible reason for this discrepancy is that the lithium production cross sections that we employ in our propagation code are incorrect. However, from the compilation of the production cross sections \citep{2018PhRvC..98c4611G}, one can see that the major production channels are fragmentation of carbon and oxygen, $^{12}$C, $^{16}$O + $p \to ^{6,7}$Li, that have been each measured in several different experiments. Even though they are not measured perfectly, each of them is contributing 12\%-14\% and thus a 20\% error in one of them would correspond to only 2\%-3\% of total lithium production. Other production channels are contributing at a level of 1\%-2\% or less. It is not impossible, but rather unlikely that cross section errors are all biased on the same side leading to the observed 20\% excess.

Finally, one can see that the AMS-02 lithium data show a kink around $\sim$100 GV (Figure \ref{fig:Li_1}), while the error bars there are rather small. Such irregularity and the relatively large scattering of data points above this rigidity may be partially responsible for the excessive hardening of the lithium spectrum above $\sim$100 GV. Here we have to wait for further data release that should indicate if this kink is an instrumental artifact or a real spectral feature and also improve the statistics at high energies. Meanwhile, since the lithium excess is present even at lower energies $\ga$4 GV, we believe that the effect is real indeed, though the fitted values may change somewhat when the more accurate data become available. 

The ultimate test of the origin of lithium in CRs would be a measurement of its isotopic composition at high energies. The nova origin of primary $^7$Li would lead to the dominance of $^7$Li at high energies, while at low energies the abundances of $^6$Li and $^7$Li are about the same. However, isotopic abundances are very difficult to measure at high energies and would require a large magnetic spectrometer with a superconducting magnet in space. The outlines of such future instruments are currently discussed in the literature \citep{2019NIMPA.94462561S}, but building and launching them into space may take a couple of decades. We, therefore, should concentrate on elimination of other possibilities through new measurements of the production cross sections or reevaluation of the available data. 


\section{Discussion}

\defcitealias{2016ApJ...824...16J}{J2016}
 
It is relevant to illustrate reasons for the apparent discrepancies between the propagation parameters derived in our present analysis and the earlier Bayesian analysis \citep[][hereafter paper \citetalias{2016ApJ...824...16J}]{2016ApJ...824...16J} of propagation of two independent sets of CR species, $\bar{p}$, $p$, He (hereafter $\bar{p}$ set), on the one hand, and Be to O nuclei (hereafter Be-O set), on the other hand, that was also based on \galprop. In short, the two sets of data ($\bar{p}$ set and Be-O set) taken separately yield normalizations of the diffusion coefficient and the Alfv\'en velocities which are inconsistent with each other \citepalias[see a plot $V_{\rm Alf}$ vs.\ $D_0$ in Figure 3 in][]{2016ApJ...824...16J}, suggesting that properties of the interstellar medium in the Galaxy differ on different scales. The latter follows from a comparison of the \emph{effective} propagation distances of these species that are significantly different due to differences in the total inelastic cross sections ($\sim$40 mb for $p$ and $\bar{p}$ vs.\ $\sim$250 mb for carbon). This conclusion is somewhat contradictory to the conclusion made in the current series of our papers \citep[][and the present paper, hereafter \emph{B-series} papers]{2017ApJ...840..115B,2018ApJ...858...61B,2018ApJ...854...94B} that the same propagation parameters can be used to describe the propagation of all light CR species, including $\bar p$, and H-O nuclei.

While the properties of the interstellar medium can vary on different scales indeed, and to definitively test it, we have to wait for accurate measurements of heavier CR species with larger fragmentation cross sections, preferably such as Fe and heavier, we shall discuss here what makes the conclusions from the two analyses somewhat different. We can see, at least, three major areas which can lead to some differences in the results: (i) the approach, (ii) the sets of data used, and (iii) the treatment of the heliospheric propagation. Here we discuss them one by one. 

(i) In the \emph{B-series} papers, in each case the propagation parameters are derived using the B/C ratio assuming \emph{diffusion-convection-reacceleration} model. In \citet{2017ApJ...840..115B} no attempt to fit the CR $\bar{p}$ spectrum was made, instead the $\bar{p}$ spectrum was calculated using the set of propagation parameters derived from the B/C ratio and fits to the spectra of protons and helium.  Besides, the $\bar p$ production cross sections used in \emph{B-series} are tuned to the modern accelerator data using the Monte Carlo event generator QGSJET-II-04m \citep{2015ApJ...803...54K,2019arXiv190405129K}. In contrast, in \citetalias{2016ApJ...824...16J}, the $\bar{p}$ set of propagation parameters was tuned to the CR $p$, He, and $\bar{p}$ spectra in \emph{diffusion-reacceleration} model \emph{excluding} all other CR species, while $\bar{p}$ production was calculated using considerably older cross section parameterizations \citep{1983JPhG....9.1289T,1983JPhG....9..227T,2002ApJ...565..280M}.

(ii) The current analysis (\emph{B-series}) uses precise AMS-02 data for all light species, such as $\bar p$, and H-O nuclei, which were not available at the time when the Bayesian analysis was ran, and thus the \citetalias{2016ApJ...824...16J} paper used mostly the ACE/CRIS, HEAO-3, and PAMELA data. Meanwhile, there are significant systematic discrepancies between different instruments (AMS-02 vs.\ HEAO-3 and PAMELA) even at the energies where the effect of the solar modulation is small, see examples in Figures~\ref{fig:BeBN_1}, \ref{fig:BeBN_2}. The CR $\bar{p}$ data also became significantly more accurate \citep[see Figure 13 in][]{2017ApJ...840..115B}.

(iii) Finally the treatment of the solar modulation became exceedingly more precise. In the \emph{B-series} we use the sophisticated HelMod code which includes the complete treatment of solar modulation mechanisms and is tuned to the data of numerous spacecraft taken at different levels of solar activity and polarities of the solar magnetic field, including measurements at different heliospheric distances (e.g., Voyager 1, 2, Pioneer 10, 11) and outside of the ecliptic plane (Ulysses), see Section~\ref{Sect::Helmod} for more details. Meanwhile, the calculations in \citetalias{2016ApJ...824...16J} paper used rather simplified the force-field approximation with a single parameter. To accommodate to the unknown systematic uncertainties of different instruments and to a priori unknown levels of the solar modulation, a total of 8 nuisance and rescaling parameters were used. Meanwhile, a proper treatment of the heliospheric propagation in \citetalias{2016ApJ...824...16J} would be unfeasible given the already heavy computation load. 

Nevertheless, the parameters derived in \citetalias{2016ApJ...824...16J} paper  for the Be-O set look quite consistent with those derived in the current \emph{B-series}. The posterior normalizations of the diffusion coefficient in the two sets of data are $D_0=6.102\pm1.662$ ($\bar{p}$ set) and $9.030\pm1.610$ (Be-O set) in units of $10^{28}$ cm$^{2}$ s$^{-1}$ \citepalias[Table 4 in][]{2016ApJ...824...16J} vs.\ $4.3\pm0.7$ (this paper, Table~\ref{tbl-prop}) in the same units and at the same normalization rigidity. The former two values look quite reasonable given that the posterior halo sizes are considerably larger in \citetalias{2016ApJ...824...16J}, $z_h=10.358\pm4.861$ kpc ($\bar{p}$ set) and $10.351\pm4.202$ kpc (Be-O set) vs.\ $4.0\pm0.6$ kpc in this paper, and so the ratios $D_0/z_h\approx 0.9\pm 0.5$ \citepalias[Be-O set,][]{2016ApJ...824...16J} and $D_0/z_h\approx 1.1\pm0.3$ (\emph{B-series}) in units of $10^{28}$ cm$^{2}$ s$^{-1}$ kpc$^{-1}$ are consistent.\footnote{Note that the ratio $D_0/z_h$ is an approximate invariant for a given propagation model \citep[for more details see, e.g.,][]{1990acr..book.....B,2007ARNPS..57..285S}.} In the case of the $\bar{p}$ set \citepalias{2016ApJ...824...16J}, $D_0/z_h\approx 0.6\pm 0.4$ in the same units, i.e.\ still consistent with other ratios. Besides, the \citetalias{2016ApJ...824...16J} calculations do not include the convection, which implies some intrinsic difference with the present analysis. 

The indices of the rigidity dependence of the diffusion coefficient are also consistent, $0.461\pm0.065$ ($\bar{p}$ set) and $0.380\pm0.018$ (Be-O set) vs.\ $0.415\pm0.025$ in this paper, and so as the Alv\'en velocity derived for the Be-O set, $V_{\rm Alf} = 30.017\pm2.461$ km s$^{-1}$ vs.\ $30\pm3$ km s$^{-1}$ (\emph{B-series}). The significant deviation is a relatively small Alfv\'en velocity, $8.970\pm1.244$ km s$^{-1}$, derived for the $\bar{p}$ set. An obvious explanation of this difference is that due to the characteristic shape of the CR $\bar{p}$ spectrum that is suppressed at low energies, the fit prefers a weak reacceleration \citep[see also][]{2002ApJ...565..280M}. If a new fit to the $\bar{p}$ data is attempted, it would likely yield the values similar to those obtained for $\bar{p}$ set in \citetalias{2016ApJ...824...16J}. Meanwhile, for a correct interpretation of the fit results, the accuracy of the used $\bar{p}$ production cross sections should be comparable to or exceed the accuracy of the CR data, and new accelerator measurements of the $\bar{p}$ production in the whole relevant energy range would be very desirable here.

In any case, it is important to repeat such types of analyses when accurate data for heavier nuclei become available in order to see if the propagation parameters derived from fits to different CR species are consistent or systematically different.

\section{Conclusion}

The paper presents our results on the analysis of the Galactic propagation and heliospheric transport of the secondary species, lithium, beryllium, boron, and nitrogen that has a considerable secondary contribution. The derived LIS allow a consistent reproduction of all available data from different instruments, while also indicate likely considerable instrumental systematic uncertainties (e.g., HEAO-3 vs.\ AMS-02). In the Appendix we provide analytical fits to the calculated LIS of lithium, beryllium, boron, and nitrogen for the energy range from 350 MV -- 50 TV, Eqs.~(\ref{eq:Li})-(\ref{eq:N}), as well as fully numerical output tables of the LIS from GALPROP runs, Tables \ref{Tbl-LithiumLIS}-\ref{Tbl-NitrogenLIS}. The fits are tuned to match the GALPROP-calculated LIS within 1\%--5\% over five orders of magnitude in rigidity including the spectral flattening at high energies. The numerical values in the fits and tables correspond to the {\it I}-scenario which has more free parameters and, therefore, the fits to the data are more accurate, see Figures~\ref{fig:AMS02_Break_2}, \ref{fig:AMS02_Break_1}. The search for the analytical expressions of the fit functions -- using the same algorithm as described in \citet{2017ApJ...840..115B} -- was guided by the advanced MCMC fitting procedure Eureqa\footnote{http://www.nutonian.com/products/eureqa/}.

Contrary to the mostly primary species, whose injection spectra could be adjusted to match the observations and thus hide the model uncertainties associated with the cross sections, assumed source and gas distributions, and other input parameters, the secondary species are not that forgiving. Instead, in the case of secondary nuclei, all errors associated with the description of propagation of primary species are popping up, even though all models are tuned to the observed secondary-to-primary nuclei ratio, typically B/C.

Therefore, the obtained overall agreement in the description of the spectra of lithium, beryllium, boron, and nitrogen in the framework of the same model that was applied to CR protons, He, C, and O is quite spectacular. The systematic overproduction of beryllium at low energies and a deficit of lithium at high energies cannot be cured through the adjustment of other parameters and, therefore, have to be taken seriously. In the case of beryllium, the most likely reason for the discrepancy are the uncertainties in the total inelastic cross sections of beryllium isotopes whose uncertainty is comparable with the value of the observed discrepancy between model predictions and CR observations and is most pronounced in the energy range below $\sim$10 GeV/n. The remarkable excess of lithium at high energies $\ga$4 GV is likely to be of a different origin. It is observed in the energy range where the solar modulation is moderate or negligible, while all involved cross sections become energy-independent. It is, therefore, likely that we see a signature of a new process. 

Recent observations of the $^7$Be lines in the spectra of novae implies that primary $^7$Li should also be present in the ejecta. The peculiar injection spectrum of primary $^7$Li as derived from the fit to the data (Table \ref{tbl-inject}) may be an indication of its non-standard origin. In the injection spectrum of $^7$Li, one can distinguish two different components: the low-energy one that has a steep spectrum between $\sim$12 GV and 350 GV with index of $2.7$, and the flat-spectrum high-energy component with index $\sim$1.9. The value of the index break $0.8\pm0.06$ at $R_2=355\pm15$ GV is the largest among the species shown in Table \ref{tbl-inject}. The low-energy part can thus be attributed to the acceleration in the nova shock, while the flat high-energy part with a universal injection index of $2$ may come from acceleration of the ejected primary lithium by a SNR shell. Such an interpretation, if correct indeed, provides a remarkable insight into the physics of primordial and stellar nucleosynthesis, and may dramatically change our understanding of the origin of CR species. One can notice that the break in the primary lithium injection spectrum $R_2$ is at about the same rigidity as for other species. This could be a chance coincidence due to the uncertainties in the lithium data at and above the break rigidity.  Meanwhile, evaluation of uncertainties associated with the primary lithium component may need a dedicated study. Though such a possibility is very exciting, we cannot rule out other possibilities for the observed discrepancy just yet. 

Interestingly, the injection ({\it I}) and propagation ({\it P}) scenarios discussed in Sections \ref{scenarios} and \ref{Sect:Results_2} provide fits of the existing data of a comparable quality. Still the {\it P}-scenario looks preferable as it does not require individual breaks in the injection spectra of all CR species, but rather one universal break in the diffusion coefficient. A better agreement with the CR anisotropy measurements is also an advantage of the {\it P}-scenario. Meanwhile, the described lithium excess is observed in both scenarios, so it is not a feature associated with one particular propagation model.  

Our continuing studies of the LIS of various CR species in the combined framework of two propagation codes, \galprop{} and \helmod{}, show that it is possible to make a self-consistent model of CR propagation in the Galaxy and the heliosphere. Meanwhile, the more and more accurate CR data available in the last decade uncover new effects and shed new light on the origin of CR species in the energy range that is deemed as well-studied. The self-consistent approach that was one of the cornerstones in the development of the \galprop{} propagation code is the key to discover such new features as it does not allow much freedom in fitting particular datasets, while using a self-consistent approach for heliospheric propagation as realized in the \helmod{} code ensures that all observed features are real indeed.

\vspace{-1\baselineskip}
\acknowledgements
Special thanks to Pavol Bobik, Giuliano Boella, Karel Kudela, Marian Putis, and Mario Zannoni for their continuous support of the \helmod{} project and many useful suggestions. This work is supported by ASI (Agenzia Spaziale Italiana) through a contract ASI-INFN I/002/13/0 and by ESA (European Space Agency) through a contract 4000116146/16/NL/HK. Igor Moskalenko and Troy Porter acknowledge support from NASA Grant No.~NNX17AB48G. We thank the ACE CRIS instrument team and the ACE Science Center for providing the ACE data.

\newpage

\bibliography{bibliography}

\appendix
\section{Supplementary Material}\label{app:SupMat}

Here we provide analytical fits to the calculated LIS of lithium, beryllium, boron, and nitrogen for the energy range from 350 MV -- 50 TV, Eqs.~(\ref{eq:Li})-(\ref{eq:N}), as well as fully numerical output tables of the LIS from GALPROP, Tables \ref{Tbl-LithiumLIS}-\ref{Tbl-NitrogenLIS}. The fits are tuned to match the GALPROP-calculated LIS within 1\%--5\% over five orders of magnitude in rigidity including the spectral flattening at high energies:
\begin{align}
\label{eq:Li}
F_{\rm Li}& (R) \times R^{2.7} = \\
&\begin{cases}
\displaystyle \frac{104.41 R^{-0.24}}{ 4.50 + 117.73 R^{-3.60}+21.45 R^{-6.72}},	  &R\le 2.7\, {\rm GV}, \smallskip\\
\displaystyle 3.11 -\frac{46.22}{R} + \frac{2509.17}{92 + 4.02 R} - 1.41\times 10^{-5}R + 4.73\times 10^{-11}R^2  + (0.032 - 6.74\times 10^{-14} R^2) \sqrt{R},  &R> 2.7\, {\rm GV},
\nonumber
\end{cases}\\[8pt]
\label{eq:Be}
F_{\rm Be}& (R) \times R^{2.7} = \\
& \begin{cases}
\displaystyle \frac{167.10 R^{-0.25}}{ 15.34 + 313.32 R^{-3.61}+49.18 R^{-6.48}},  &R\le 2.3\, {\rm GV}, \smallskip\\
\displaystyle 0.067 - \left(4.12\times 10^{-5}  -3.21\times 10^{-6} \ln R + 0.18 \times 0.90^R\right) R + \left(0.99 - 1.59 \times 0.32^R \right) e^{\frac{3.16 \ln R} {\sqrt{0.20 + R}}} ,  &R> 2.3\, {\rm GV},
\nonumber
\end{cases}\\[8pt]
\label{eq:B}
F_{\rm B}& (R) \times R^{2.7} = \\
& \begin{cases}
\displaystyle \frac{11.47 R^{-0.25}}{ 0.327 + 7.50 R^{-3.67}+1.21 R^{-6.73}},  &R\le 3\, {\rm GV}, \smallskip\\
\displaystyle 6.90 + \frac{946.97}{61.11 + R}- \frac{606.57 + 37.17\widetilde{R}}{R} + \frac{1004.79\widetilde{R}}{R + R\widetilde{R}} + 0.00017 R - 9.79\times 10^{-10}R^2 - 0.0465\widetilde{R},  &R> 3\, {\rm GV},
\nonumber
\end{cases}\\[8pt]
\label{eq:N}
F_{\rm N}& (R) \times R^{2.7} = \\
& \begin{cases}
\displaystyle \frac{ 75.12 R^{-0.16}}{2.69 + 29.18 R^{-2.75} +21.81 R^{-5.25}},  &R\le 3.6\, {\rm GV}, \smallskip\\
\displaystyle 281.08 + \frac{65}{1.66 + 1.0072 R} + 0.0017 R - 0.87 \sqrt R + (24.5 - 4.19\times 10^{-5} R) \sqrt[4]R - 79.07\ln (21.52 \sqrt[4]R),  &R> 3.6\, {\rm GV},
\nonumber
\end{cases}
\end{align}
where $\widetilde{R}=\sqrt{1.19 + R}$. The numerical values in the fits and tables correspond to the {\it I}-scenario which has more free parameters and, therefore, the fits to the data are more accurate, see Figures~\ref{fig:AMS02_Break_2}, \ref{fig:AMS02_Break_1}. 


\begin{deluxetable}{cccccccccc}[p]
	\tablecolumns{10}
	\tablewidth{0mm}
	\tablecaption{Lithium LIS\label{Tbl-LithiumLIS}}
	\tablehead{
		\colhead{Rigidity} & \colhead{Differential} &
		\colhead{Rigidity} & \colhead{Differential} &
		\colhead{Rigidity} & \colhead{Differential} &
		\colhead{Rigidity} & \colhead{Differential} &
		\colhead{Rigidity} & \colhead{Differential}
		\\
		\colhead{GV} & \colhead{Intensity\tablenotemark{a}} &
		\colhead{GV} & \colhead{Intensity\tablenotemark{a}} &
		\colhead{GV} & \colhead{Intensity\tablenotemark{a}} &
		\colhead{GV} & \colhead{Intensity\tablenotemark{a}} &
		\colhead{GV} & \colhead{Intensity\tablenotemark{a}} 
	}
\startdata
1.007e-01 & 3.173e-03 & 6.359e-01 & 3.767e-01 & 5.303e+00 & 1.837e-01 & 1.552e+02 & 8.331e-06 & 5.976e+03 & 3.570e-10\\
1.057e-01 & 3.634e-03 & 6.680e-01 & 4.209e-01 & 5.691e+00 & 1.541e-01 & 1.706e+02 & 6.196e-06 & 6.581e+03 & 2.802e-10\\
1.109e-01 & 4.106e-03 & 7.018e-01 & 4.656e-01 & 6.113e+00 & 1.285e-01 & 1.877e+02 & 4.608e-06 & 7.247e+03 & 2.201e-10\\
1.164e-01 & 4.638e-03 & 7.374e-01 & 5.088e-01 & 6.575e+00 & 1.065e-01 & 2.065e+02 & 3.429e-06 & 7.981e+03 & 1.730e-10\\
1.222e-01 & 5.241e-03 & 7.749e-01 & 5.489e-01 & 7.080e+00 & 8.777e-02 & 2.272e+02 & 2.553e-06 & 8.789e+03 & 1.361e-10\\
1.282e-01 & 5.923e-03 & 8.144e-01 & 5.856e-01 & 7.633e+00 & 7.201e-02 & 2.500e+02 & 1.903e-06 & 9.678e+03 & 1.071e-10\\
1.346e-01 & 6.695e-03 & 8.561e-01 & 6.181e-01 & 8.238e+00 & 5.881e-02 & 2.751e+02 & 1.421e-06 & 1.066e+04 & 8.438e-11\\
1.412e-01 & 7.569e-03 & 9.000e-01 & 6.481e-01 & 8.900e+00 & 4.784e-02 & 3.027e+02 & 1.063e-06 & 1.174e+04 & 6.651e-11\\
1.482e-01 & 8.559e-03 & 9.463e-01 & 6.776e-01 & 9.627e+00 & 3.875e-02 & 3.331e+02 & 7.976e-07 & 1.293e+04 & 5.245e-11\\
1.555e-01 & 9.681e-03 & 9.953e-01 & 7.088e-01 & 1.042e+01 & 3.124e-02 & 3.666e+02 & 6.007e-07 & 1.423e+04 & 4.139e-11\\
1.632e-01 & 1.095e-02 & 1.047e+00 & 7.434e-01 & 1.130e+01 & 2.505e-02 & 4.035e+02 & 4.542e-07 & 1.567e+04 & 3.268e-11\\
1.713e-01 & 1.239e-02 & 1.102e+00 & 7.795e-01 & 1.226e+01 & 1.996e-02 & 4.442e+02 & 3.447e-07 & 1.726e+04 & 2.582e-11\\
1.798e-01 & 1.402e-02 & 1.159e+00 & 8.161e-01 & 1.332e+01 & 1.579e-02 & 4.889e+02 & 2.624e-07 & 1.901e+04 & 2.041e-11\\
1.887e-01 & 1.587e-02 & 1.220e+00 & 8.517e-01 & 1.448e+01 & 1.241e-02 & 5.382e+02 & 2.004e-07 & 2.093e+04 & 1.614e-11\\
1.980e-01 & 1.797e-02 & 1.285e+00 & 8.850e-01 & 1.575e+01 & 9.691e-03 & 5.925e+02 & 1.534e-07 & 2.305e+04 & 1.278e-11\\
2.078e-01 & 2.035e-02 & 1.354e+00 & 9.151e-01 & 1.715e+01 & 7.525e-03 & 6.522e+02 & 1.176e-07 & 2.539e+04 & 1.011e-11\\
2.181e-01 & 2.304e-02 & 1.427e+00 & 9.407e-01 & 1.869e+01 & 5.814e-03 & 7.180e+02 & 9.035e-08 & 2.796e+04 & 8.012e-12\\
2.289e-01 & 2.610e-02 & 1.504e+00 & 9.608e-01 & 2.039e+01 & 4.473e-03 & 7.905e+02 & 6.951e-08 & 3.079e+04 & 6.349e-12\\
2.403e-01 & 2.957e-02 & 1.586e+00 & 9.745e-01 & 2.226e+01 & 3.428e-03 & 8.703e+02 & 5.355e-08 & 3.390e+04 & 5.034e-12\\
2.522e-01 & 3.351e-02 & 1.673e+00 & 9.810e-01 & 2.431e+01 & 2.619e-03 & 9.582e+02 & 4.131e-08 & 3.733e+04 & 3.993e-12\\
2.647e-01 & 3.799e-02 & 1.766e+00 & 9.795e-01 & 2.657e+01 & 1.996e-03 & 1.055e+03 & 3.190e-08 & 4.111e+04 & 3.168e-12\\
2.778e-01 & 4.307e-02 & 1.865e+00 & 9.695e-01 & 2.905e+01 & 1.517e-03 & 1.162e+03 & 2.466e-08 & 4.528e+04 & 2.515e-12\\
2.916e-01 & 4.886e-02 & 1.971e+00 & 9.508e-01 & 3.179e+01 & 1.150e-03 & 1.279e+03 & 1.909e-08 & 4.986e+04 & 1.997e-12\\
3.061e-01 & 5.544e-02 & 2.084e+00 & 9.236e-01 & 3.480e+01 & 8.701e-04 & 1.408e+03 & 1.479e-08 & 5.491e+04 & 1.587e-12\\
3.213e-01 & 6.293e-02 & 2.204e+00 & 8.881e-01 & 3.812e+01 & 6.571e-04 & 1.551e+03 & 1.147e-08 & 6.047e+04 & 1.261e-12\\
3.372e-01 & 7.147e-02 & 2.334e+00 & 8.453e-01 & 4.177e+01 & 4.953e-04 & 1.707e+03 & 8.906e-09 & 6.659e+04 & 1.002e-12\\
3.540e-01 & 8.121e-02 & 2.472e+00 & 7.959e-01 & 4.579e+01 & 3.727e-04 & 1.880e+03 & 6.919e-09 & 7.333e+04 & 7.972e-13\\
3.716e-01 & 9.233e-02 & 2.621e+00 & 7.414e-01 & 5.022e+01 & 2.800e-04 & 2.070e+03 & 5.380e-09 & 8.076e+04 & 6.341e-13\\
3.901e-01 & 1.050e-01 & 2.781e+00 & 6.837e-01 & 5.509e+01 & 2.100e-04 & 2.280e+03 & 4.187e-09 & 8.893e+04 & 5.046e-13\\
4.096e-01 & 1.196e-01 & 2.954e+00 & 6.243e-01 & 6.046e+01 & 1.573e-04 & 2.510e+03 & 3.261e-09 & 9.794e+04 & 4.016e-13\\
4.300e-01 & 1.363e-01 & 3.139e+00 & 5.645e-01 & 6.637e+01 & 1.177e-04 & 2.764e+03 & 2.543e-09 & 1.079e+05 & 3.197e-13\\
4.514e-01 & 1.554e-01 & 3.340e+00 & 5.060e-01 & 7.287e+01 & 8.793e-05 & 3.044e+03 & 1.984e-09 & 1.188e+05 & 2.546e-13\\
4.740e-01 & 1.771e-01 & 3.557e+00 & 4.495e-01 & 8.004e+01 & 6.565e-05 & 3.352e+03 & 1.549e-09 & 1.308e+05 & 2.028e-13\\
4.977e-01 & 2.017e-01 & 3.791e+00 & 3.956e-01 & 8.792e+01 & 4.897e-05 & 3.691e+03 & 1.211e-09 & 1.440e+05 & 1.616e-13\\
5.226e-01 & 2.295e-01 & 4.046e+00 & 3.451e-01 & 9.661e+01 & 3.650e-05 & 4.064e+03 & 9.470e-10 & 1.586e+05 & 1.287e-13\\
5.488e-01 & 2.609e-01 & 4.322e+00 & 2.983e-01 & 1.062e+02 & 2.719e-05 & 4.476e+03 & 7.412e-10 & 1.747e+05 & 1.026e-13\\
5.764e-01 & 2.961e-01 & 4.622e+00 & 2.558e-01 & 1.167e+02 & 2.024e-05 & 4.928e+03 & 5.806e-10 & 1.924e+05 & 8.176e-14\\
6.054e-01 & 3.349e-01 & 4.948e+00 & 2.176e-01 & 1.283e+02 & 1.506e-05 & 5.427e+03 & 4.551e-10 & 2.118e+05 & 6.486e-14
\enddata
	\tablenotetext{a}{Differential Intensity units: (m$^2$ s sr GV)$^{-1}$.}
\end{deluxetable}

\begin{deluxetable}{cccccccccc}[p]
	\tablecolumns{10}
	\tablewidth{0mm}
	\tablecaption{Beryllium LIS\label{Tbl-BerylliumLIS}}
	\tablehead{
		\colhead{Rigidity} & \colhead{Differential} &
		\colhead{Rigidity} & \colhead{Differential} &
		\colhead{Rigidity} & \colhead{Differential} &
		\colhead{Rigidity} & \colhead{Differential} &
		\colhead{Rigidity} & \colhead{Differential}
		\\
		\colhead{GV} & \colhead{Intensity\tablenotemark{a}} &
		\colhead{GV} & \colhead{Intensity\tablenotemark{a}} &
		\colhead{GV} & \colhead{Intensity\tablenotemark{a}} &
		\colhead{GV} & \colhead{Intensity\tablenotemark{a}} &
		\colhead{GV} & \colhead{Intensity\tablenotemark{a}} 
	}
\startdata
7.555e-02 & 6.565e-04 & 4.769e-01 & 1.412e-01 & 3.977e+00 & 1.889e-01 & 1.164e+02 & 1.078e-05 & 4.482e+03 & 2.078e-10\\
7.929e-02 & 8.939e-04 & 5.010e-01 & 1.581e-01 & 4.268e+00 & 1.624e-01 & 1.280e+02 & 8.021e-06 & 4.936e+03 & 1.570e-10\\
8.321e-02 & 1.116e-03 & 5.264e-01 & 1.759e-01 & 4.585e+00 & 1.383e-01 & 1.408e+02 & 5.968e-06 & 5.436e+03 & 1.186e-10\\
8.732e-02 & 1.340e-03 & 5.531e-01 & 1.944e-01 & 4.931e+00 & 1.167e-01 & 1.549e+02 & 4.440e-06 & 5.986e+03 & 8.959e-11\\
9.164e-02 & 1.570e-03 & 5.812e-01 & 2.138e-01 & 5.310e+00 & 9.759e-02 & 1.704e+02 & 3.303e-06 & 6.592e+03 & 6.767e-11\\
9.617e-02 & 1.812e-03 & 6.108e-01 & 2.340e-01 & 5.724e+00 & 8.101e-02 & 1.875e+02 & 2.458e-06 & 7.259e+03 & 5.111e-11\\
1.009e-01 & 2.073e-03 & 6.421e-01 & 2.551e-01 & 6.178e+00 & 6.673e-02 & 2.063e+02 & 1.830e-06 & 7.994e+03 & 3.859e-11\\
1.059e-01 & 2.361e-03 & 6.750e-01 & 2.768e-01 & 6.675e+00 & 5.458e-02 & 2.270e+02 & 1.363e-06 & 8.803e+03 & 2.914e-11\\
1.112e-01 & 2.688e-03 & 7.098e-01 & 2.991e-01 & 7.220e+00 & 4.437e-02 & 2.498e+02 & 1.016e-06 & 9.694e+03 & 2.200e-11\\
1.167e-01 & 3.062e-03 & 7.465e-01 & 3.213e-01 & 7.818e+00 & 3.585e-02 & 2.750e+02 & 7.576e-07 & 1.068e+04 & 1.661e-11\\
1.224e-01 & 3.487e-03 & 7.852e-01 & 3.429e-01 & 8.475e+00 & 2.878e-02 & 3.026e+02 & 5.657e-07 & 1.176e+04 & 1.254e-11\\
1.285e-01 & 3.972e-03 & 8.262e-01 & 3.637e-01 & 9.195e+00 & 2.294e-02 & 3.331e+02 & 4.228e-07 & 1.295e+04 & 9.466e-12\\
1.348e-01 & 4.526e-03 & 8.695e-01 & 3.843e-01 & 9.987e+00 & 1.821e-02 & 3.667e+02 & 3.163e-07 & 1.426e+04 & 7.145e-12\\
1.415e-01 & 5.158e-03 & 9.153e-01 & 4.049e-01 & 1.086e+01 & 1.436e-02 & 4.036e+02 & 2.370e-07 & 1.570e+04 & 5.393e-12\\
1.485e-01 & 5.879e-03 & 9.639e-01 & 4.252e-01 & 1.181e+01 & 1.125e-02 & 4.443e+02 & 1.777e-07 & 1.729e+04 & 4.070e-12\\
1.559e-01 & 6.702e-03 & 1.015e+00 & 4.452e-01 & 1.286e+01 & 8.774e-03 & 4.892e+02 & 1.334e-07 & 1.904e+04 & 3.071e-12\\
1.636e-01 & 7.641e-03 & 1.070e+00 & 4.648e-01 & 1.402e+01 & 6.821e-03 & 5.385e+02 & 1.002e-07 & 2.097e+04 & 2.318e-12\\
1.717e-01 & 8.714e-03 & 1.128e+00 & 4.837e-01 & 1.529e+01 & 5.286e-03 & 5.929e+02 & 7.541e-08 & 2.309e+04 & 1.749e-12\\
1.802e-01 & 9.938e-03 & 1.189e+00 & 5.020e-01 & 1.669e+01 & 4.083e-03 & 6.527e+02 & 5.677e-08 & 2.543e+04 & 1.319e-12\\
1.891e-01 & 1.134e-02 & 1.255e+00 & 5.191e-01 & 1.823e+01 & 3.143e-03 & 7.187e+02 & 4.277e-08 & 2.800e+04 & 9.955e-13\\
1.985e-01 & 1.293e-02 & 1.324e+00 & 5.340e-01 & 1.993e+01 & 2.412e-03 & 7.913e+02 & 3.224e-08 & 3.084e+04 & 7.511e-13\\
2.084e-01 & 1.476e-02 & 1.399e+00 & 5.458e-01 & 2.179e+01 & 1.845e-03 & 8.712e+02 & 2.432e-08 & 3.396e+04 & 5.666e-13\\
2.187e-01 & 1.685e-02 & 1.478e+00 & 5.538e-01 & 2.384e+01 & 1.408e-03 & 9.593e+02 & 1.836e-08 & 3.740e+04 & 4.275e-13\\
2.296e-01 & 1.923e-02 & 1.563e+00 & 5.576e-01 & 2.610e+01 & 1.071e-03 & 1.056e+03 & 1.386e-08 & 4.118e+04 & 3.225e-13\\
2.410e-01 & 2.196e-02 & 1.653e+00 & 5.566e-01 & 2.859e+01 & 8.132e-04 & 1.163e+03 & 1.047e-08 & 4.535e+04 & 2.433e-13\\
2.529e-01 & 2.508e-02 & 1.750e+00 & 5.507e-01 & 3.133e+01 & 6.157e-04 & 1.281e+03 & 7.909e-09 & 4.994e+04 & 1.835e-13\\
2.655e-01 & 2.864e-02 & 1.854e+00 & 5.396e-01 & 3.434e+01 & 4.650e-04 & 1.410e+03 & 5.977e-09 & 5.500e+04 & 1.384e-13\\
2.787e-01 & 3.273e-02 & 1.966e+00 & 5.234e-01 & 3.766e+01 & 3.505e-04 & 1.553e+03 & 4.517e-09 & 6.057e+04 & 1.044e-13\\
2.926e-01 & 3.741e-02 & 2.086e+00 & 5.025e-01 & 4.132e+01 & 2.636e-04 & 1.710e+03 & 3.414e-09 & 6.670e+04 & 7.875e-14\\
3.072e-01 & 4.277e-02 & 2.215e+00 & 4.773e-01 & 4.534e+01 & 1.980e-04 & 1.883e+03 & 2.580e-09 & 7.345e+04 & 5.940e-14\\
3.225e-01 & 4.891e-02 & 2.354e+00 & 4.484e-01 & 4.977e+01 & 1.485e-04 & 2.073e+03 & 1.951e-09 & 8.089e+04 & 4.480e-14\\
3.386e-01 & 5.594e-02 & 2.505e+00 & 4.168e-01 & 5.465e+01 & 1.114e-04 & 2.283e+03 & 1.474e-09 & 8.908e+04 & 3.379e-14\\
3.555e-01 & 6.400e-02 & 2.668e+00 & 3.833e-01 & 6.003e+01 & 8.343e-05 & 2.514e+03 & 1.115e-09 & 9.810e+04 & 2.549e-14\\
3.733e-01 & 7.326e-02 & 2.844e+00 & 3.490e-01 & 6.594e+01 & 6.243e-05 & 2.768e+03 & 8.427e-10 & 1.080e+05 & 1.922e-14\\
3.920e-01 & 8.396e-02 & 3.034e+00 & 3.147e-01 & 7.246e+01 & 4.669e-05 & 3.048e+03 & 6.370e-10 & 1.190e+05 & 1.450e-14\\
4.116e-01 & 9.622e-02 & 3.241e+00 & 2.810e-01 & 7.963e+01 & 3.488e-05 & 3.357e+03 & 4.815e-10 & 1.310e+05 & 1.094e-14\\
4.323e-01 & 1.101e-01 & 3.466e+00 & 2.485e-01 & 8.753e+01 & 2.604e-05 & 3.696e+03 & 3.639e-10 & 1.443e+05 & 8.248e-15\\
4.540e-01 & 1.251e-01 & 3.711e+00 & 2.177e-01 & 9.623e+01 & 1.942e-05 & 4.070e+03 & 2.750e-10 & 1.589e+05 & 6.218e-15
\enddata
	\tablenotetext{a}{Differential Intensity units: (m$^2$ s sr GV)$^{-1}$.}
\end{deluxetable}

\begin{deluxetable}{cccccccccc}[p]
	\tablecolumns{10}
	\tablewidth{0mm}
	\tablecaption{Boron LIS\label{Tbl-BoronLIS}}
	\tablehead{
		\colhead{Rigidity} & \colhead{Differential} &
		\colhead{Rigidity} & \colhead{Differential} &
		\colhead{Rigidity} & \colhead{Differential} &
		\colhead{Rigidity} & \colhead{Differential} &
		\colhead{Rigidity} & \colhead{Differential}
		\\
		\colhead{GV} & \colhead{Intensity\tablenotemark{a}} &
		\colhead{GV} & \colhead{Intensity\tablenotemark{a}} &
		\colhead{GV} & \colhead{Intensity\tablenotemark{a}} &
		\colhead{GV} & \colhead{Intensity\tablenotemark{a}} &
		\colhead{GV} & \colhead{Intensity\tablenotemark{a}} 
	}
	\startdata
	9.498e-02 & 4.269e-03 & 5.996e-01 & 5.957e-01 & 5.000e+00 & 3.198e-01 & 1.463e+02 & 1.438e-05 & 5.635e+03 & 3.042e-10\\
	9.968e-02 & 4.902e-03 & 6.298e-01 & 6.625e-01 & 5.365e+00 & 2.671e-01 & 1.609e+02 & 1.072e-05 & 6.205e+03 & 2.300e-10\\
	1.046e-01 & 5.571e-03 & 6.617e-01 & 7.306e-01 & 5.764e+00 & 2.216e-01 & 1.770e+02 & 7.987e-06 & 6.833e+03 & 1.739e-10\\
	1.098e-01 & 6.332e-03 & 6.953e-01 & 7.991e-01 & 6.199e+00 & 1.827e-01 & 1.947e+02 & 5.954e-06 & 7.525e+03 & 1.315e-10\\
	1.152e-01 & 7.199e-03 & 7.306e-01 & 8.673e-01 & 6.676e+00 & 1.498e-01 & 2.142e+02 & 4.441e-06 & 8.287e+03 & 9.938e-11\\
	1.209e-01 & 8.186e-03 & 7.679e-01 & 9.348e-01 & 7.196e+00 & 1.221e-01 & 2.357e+02 & 3.314e-06 & 9.125e+03 & 7.511e-11\\
	1.269e-01 & 9.310e-03 & 8.071e-01 & 1.001e+00 & 7.767e+00 & 9.895e-02 & 2.593e+02 & 2.475e-06 & 1.005e+04 & 5.676e-11\\
	1.332e-01 & 1.059e-02 & 8.486e-01 & 1.065e+00 & 8.392e+00 & 7.976e-02 & 2.854e+02 & 1.850e-06 & 1.107e+04 & 4.289e-11\\
	1.397e-01 & 1.205e-02 & 8.923e-01 & 1.128e+00 & 9.077e+00 & 6.391e-02 & 3.141e+02 & 1.384e-06 & 1.219e+04 & 3.241e-11\\
	1.467e-01 & 1.372e-02 & 9.384e-01 & 1.190e+00 & 9.829e+00 & 5.096e-02 & 3.457e+02 & 1.036e-06 & 1.342e+04 & 2.448e-11\\
	1.539e-01 & 1.561e-02 & 9.871e-01 & 1.250e+00 & 1.065e+01 & 4.046e-02 & 3.805e+02 & 7.770e-07 & 1.478e+04 & 1.849e-11\\
	1.615e-01 & 1.778e-02 & 1.039e+00 & 1.308e+00 & 1.156e+01 & 3.199e-02 & 4.188e+02 & 5.832e-07 & 1.627e+04 & 1.397e-11\\
	1.695e-01 & 2.024e-02 & 1.093e+00 & 1.365e+00 & 1.256e+01 & 2.518e-02 & 4.610e+02 & 4.383e-07 & 1.792e+04 & 1.055e-11\\
	1.779e-01 & 2.306e-02 & 1.151e+00 & 1.420e+00 & 1.365e+01 & 1.974e-02 & 5.074e+02 & 3.297e-07 & 1.974e+04 & 7.969e-12\\
	1.867e-01 & 2.627e-02 & 1.212e+00 & 1.474e+00 & 1.485e+01 & 1.540e-02 & 5.586e+02 & 2.483e-07 & 2.173e+04 & 6.018e-12\\
	1.960e-01 & 2.993e-02 & 1.276e+00 & 1.525e+00 & 1.617e+01 & 1.198e-02 & 6.149e+02 & 1.871e-07 & 2.394e+04 & 4.544e-12\\
	2.057e-01 & 3.410e-02 & 1.345e+00 & 1.573e+00 & 1.763e+01 & 9.281e-03 & 6.770e+02 & 1.411e-07 & 2.636e+04 & 3.431e-12\\
	2.158e-01 & 3.887e-02 & 1.418e+00 & 1.616e+00 & 1.923e+01 & 7.169e-03 & 7.453e+02 & 1.065e-07 & 2.903e+04 & 2.591e-12\\
	2.265e-01 & 4.431e-02 & 1.495e+00 & 1.650e+00 & 2.098e+01 & 5.521e-03 & 8.206e+02 & 8.047e-08 & 3.197e+04 & 1.956e-12\\
	2.378e-01 & 5.051e-02 & 1.577e+00 & 1.674e+00 & 2.292e+01 & 4.239e-03 & 9.035e+02 & 6.081e-08 & 3.520e+04 & 1.477e-12\\
	2.496e-01 & 5.760e-02 & 1.665e+00 & 1.685e+00 & 2.505e+01 & 3.245e-03 & 9.947e+02 & 4.597e-08 & 3.877e+04 & 1.115e-12\\
	2.619e-01 & 6.568e-02 & 1.758e+00 & 1.680e+00 & 2.739e+01 & 2.476e-03 & 1.095e+03 & 3.476e-08 & 4.269e+04 & 8.415e-13\\
	2.749e-01 & 7.491e-02 & 1.858e+00 & 1.659e+00 & 2.998e+01 & 1.885e-03 & 1.206e+03 & 2.630e-08 & 4.701e+04 & 6.352e-13\\
	2.886e-01 & 8.545e-02 & 1.965e+00 & 1.620e+00 & 3.282e+01 & 1.432e-03 & 1.328e+03 & 1.991e-08 & 5.177e+04 & 4.795e-13\\
	3.029e-01 & 9.750e-02 & 2.078e+00 & 1.563e+00 & 3.594e+01 & 1.085e-03 & 1.462e+03 & 1.507e-08 & 5.701e+04 & 3.619e-13\\
	3.180e-01 & 1.113e-01 & 2.200e+00 & 1.491e+00 & 3.939e+01 & 8.208e-04 & 1.610e+03 & 1.140e-08 & 6.279e+04 & 2.731e-13\\
	3.338e-01 & 1.270e-01 & 2.331e+00 & 1.405e+00 & 4.318e+01 & 6.197e-04 & 1.773e+03 & 8.631e-09 & 6.914e+04 & 2.062e-13\\
	3.504e-01 & 1.450e-01 & 2.471e+00 & 1.310e+00 & 4.735e+01 & 4.671e-04 & 1.952e+03 & 6.532e-09 & 7.614e+04 & 1.556e-13\\
	3.678e-01 & 1.656e-01 & 2.622e+00 & 1.209e+00 & 5.194e+01 & 3.515e-04 & 2.149e+03 & 4.944e-09 & 8.385e+04 & 1.174e-13\\
	3.861e-01 & 1.891e-01 & 2.785e+00 & 1.103e+00 & 5.700e+01 & 2.641e-04 & 2.367e+03 & 3.743e-09 & 9.234e+04 & 8.863e-14\\
	4.054e-01 & 2.160e-01 & 2.960e+00 & 9.977e-01 & 6.257e+01 & 1.982e-04 & 2.606e+03 & 2.833e-09 & 1.017e+05 & 6.689e-14\\
	4.256e-01 & 2.466e-01 & 3.149e+00 & 8.937e-01 & 6.871e+01 & 1.486e-04 & 2.870e+03 & 2.145e-09 & 1.120e+05 & 5.048e-14\\
	4.469e-01 & 2.815e-01 & 3.353e+00 & 7.934e-01 & 7.546e+01 & 1.113e-04 & 3.160e+03 & 1.623e-09 & 1.233e+05 & 3.809e-14\\
	4.693e-01 & 3.211e-01 & 3.575e+00 & 6.976e-01 & 8.290e+01 & 8.323e-05 & 3.480e+03 & 1.228e-09 & 1.358e+05 & 2.875e-14\\
	4.928e-01 & 3.656e-01 & 3.815e+00 & 6.076e-01 & 9.109e+01 & 6.222e-05 & 3.832e+03 & 9.294e-10 & 1.496e+05 & 2.170e-14\\
	5.175e-01 & 4.152e-01 & 4.075e+00 & 5.243e-01 & 1.001e+02 & 4.647e-05 & 4.220e+03 & 7.031e-10 & 1.647e+05 & 1.637e-14\\
	5.435e-01 & 4.704e-01 & 4.358e+00 & 4.483e-01 & 1.100e+02 & 3.469e-05 & 4.647e+03 & 5.319e-10 & 1.814e+05 & 1.235e-14\\
	5.708e-01 & 5.311e-01 & 4.665e+00 & 3.801e-01 & 1.210e+02 & 2.587e-05 & 5.117e+03 & 4.023e-10 & 1.997e+05 & 9.218e-15
	\enddata
	\tablenotetext{a}{Differential Intensity units: (m$^2$ s sr GV)$^{-1}$.}
\end{deluxetable}

\begin{deluxetable}{cccccccccc}[p]
	\tablecolumns{10}
	\tablewidth{0mm}
	\tablecaption{Nitrogen LIS\label{Tbl-NitrogenLIS}}
	\tablehead{
		\colhead{Rigidity} & \colhead{Differential} &
		\colhead{Rigidity} & \colhead{Differential} &
		\colhead{Rigidity} & \colhead{Differential} &
		\colhead{Rigidity} & \colhead{Differential} &
		\colhead{Rigidity} & \colhead{Differential}
		\\
		\colhead{GV} & \colhead{Intensity\tablenotemark{a}} &
		\colhead{GV} & \colhead{Intensity\tablenotemark{a}} &
		\colhead{GV} & \colhead{Intensity\tablenotemark{a}} &
		\colhead{GV} & \colhead{Intensity\tablenotemark{a}} &
		\colhead{GV} & \colhead{Intensity\tablenotemark{a}} 
	}
\startdata
9.252e-02 & 1.160e-02 & 5.840e-01 & 6.888e-01 & 4.870e+00 & 2.867e-01 & 1.425e+02 & 2.198e-05 & 5.489e+03 & 1.780e-09\\
9.709e-02 & 1.299e-02 & 6.135e-01 & 7.493e-01 & 5.226e+00 & 2.417e-01 & 1.567e+02 & 1.662e-05 & 6.044e+03 & 1.416e-09\\
1.019e-01 & 1.453e-02 & 6.445e-01 & 8.130e-01 & 5.614e+00 & 2.026e-01 & 1.724e+02 & 1.257e-05 & 6.656e+03 & 1.128e-09\\
1.069e-01 & 1.625e-02 & 6.772e-01 & 8.796e-01 & 6.038e+00 & 1.689e-01 & 1.896e+02 & 9.517e-06 & 7.330e+03 & 8.984e-10\\
1.122e-01 & 1.817e-02 & 7.116e-01 & 9.485e-01 & 6.502e+00 & 1.402e-01 & 2.086e+02 & 7.213e-06 & 8.071e+03 & 7.160e-10\\
1.178e-01 & 2.033e-02 & 7.479e-01 & 1.019e+00 & 7.009e+00 & 1.158e-01 & 2.296e+02 & 5.475e-06 & 8.888e+03 & 5.708e-10\\
1.236e-01 & 2.274e-02 & 7.862e-01 & 1.090e+00 & 7.565e+00 & 9.522e-02 & 2.526e+02 & 4.164e-06 & 9.788e+03 & 4.552e-10\\
1.297e-01 & 2.543e-02 & 8.265e-01 & 1.160e+00 & 8.174e+00 & 7.788e-02 & 2.780e+02 & 3.175e-06 & 1.078e+04 & 3.631e-10\\
1.361e-01 & 2.845e-02 & 8.691e-01 & 1.228e+00 & 8.841e+00 & 6.336e-02 & 3.059e+02 & 2.428e-06 & 1.187e+04 & 2.898e-10\\
1.428e-01 & 3.182e-02 & 9.140e-01 & 1.295e+00 & 9.574e+00 & 5.130e-02 & 3.367e+02 & 1.862e-06 & 1.307e+04 & 2.313e-10\\
1.499e-01 & 3.558e-02 & 9.615e-01 & 1.357e+00 & 1.038e+01 & 4.137e-02 & 3.706e+02 & 1.433e-06 & 1.439e+04 & 1.847e-10\\
1.573e-01 & 3.980e-02 & 1.012e+00 & 1.414e+00 & 1.126e+01 & 3.322e-02 & 4.079e+02 & 1.106e-06 & 1.585e+04 & 1.475e-10\\
1.651e-01 & 4.450e-02 & 1.065e+00 & 1.465e+00 & 1.223e+01 & 2.657e-02 & 4.490e+02 & 8.558e-07 & 1.746e+04 & 1.178e-10\\
1.733e-01 & 4.976e-02 & 1.121e+00 & 1.508e+00 & 1.329e+01 & 2.116e-02 & 4.943e+02 & 6.643e-07 & 1.922e+04 & 9.417e-11\\
1.819e-01 & 5.563e-02 & 1.180e+00 & 1.544e+00 & 1.447e+01 & 1.678e-02 & 5.441e+02 & 5.169e-07 & 2.117e+04 & 7.527e-11\\
1.909e-01 & 6.218e-02 & 1.243e+00 & 1.573e+00 & 1.575e+01 & 1.326e-02 & 5.990e+02 & 4.031e-07 & 2.331e+04 & 6.018e-11\\
2.003e-01 & 6.949e-02 & 1.310e+00 & 1.596e+00 & 1.717e+01 & 1.043e-02 & 6.594e+02 & 3.149e-07 & 2.567e+04 & 4.812e-11\\
2.102e-01 & 7.765e-02 & 1.381e+00 & 1.609e+00 & 1.873e+01 & 8.176e-03 & 7.260e+02 & 2.464e-07 & 2.827e+04 & 3.849e-11\\
2.207e-01 & 8.674e-02 & 1.456e+00 & 1.611e+00 & 2.044e+01 & 6.386e-03 & 7.993e+02 & 1.931e-07 & 3.114e+04 & 3.079e-11\\
2.316e-01 & 9.687e-02 & 1.536e+00 & 1.602e+00 & 2.232e+01 & 4.970e-03 & 8.800e+02 & 1.516e-07 & 3.429e+04 & 2.464e-11\\
2.431e-01 & 1.081e-01 & 1.622e+00 & 1.581e+00 & 2.440e+01 & 3.856e-03 & 9.689e+02 & 1.191e-07 & 3.776e+04 & 1.972e-11\\
2.551e-01 & 1.207e-01 & 1.713e+00 & 1.547e+00 & 2.668e+01 & 2.983e-03 & 1.067e+03 & 9.369e-08 & 4.158e+04 & 1.578e-11\\
2.678e-01 & 1.347e-01 & 1.810e+00 & 1.500e+00 & 2.920e+01 & 2.301e-03 & 1.175e+03 & 7.379e-08 & 4.579e+04 & 1.263e-11\\
2.811e-01 & 1.502e-01 & 1.913e+00 & 1.442e+00 & 3.196e+01 & 1.771e-03 & 1.293e+03 & 5.817e-08 & 5.043e+04 & 1.011e-11\\
2.950e-01 & 1.675e-01 & 2.024e+00 & 1.374e+00 & 3.501e+01 & 1.360e-03 & 1.424e+03 & 4.589e-08 & 5.553e+04 & 8.098e-12\\
3.097e-01 & 1.866e-01 & 2.143e+00 & 1.299e+00 & 3.836e+01 & 1.043e-03 & 1.568e+03 & 3.623e-08 & 6.115e+04 & 6.485e-12\\
3.251e-01 & 2.079e-01 & 2.270e+00 & 1.217e+00 & 4.205e+01 & 7.977e-04 & 1.727e+03 & 2.863e-08 & 6.735e+04 & 5.194e-12\\
3.413e-01 & 2.315e-01 & 2.407e+00 & 1.131e+00 & 4.612e+01 & 6.091e-04 & 1.901e+03 & 2.264e-08 & 7.416e+04 & 4.160e-12\\
3.583e-01 & 2.576e-01 & 2.554e+00 & 1.043e+00 & 5.059e+01 & 4.644e-04 & 2.094e+03 & 1.791e-08 & 8.167e+04 & 3.333e-12\\
3.761e-01 & 2.866e-01 & 2.712e+00 & 9.527e-01 & 5.552e+01 & 3.535e-04 & 2.305e+03 & 1.418e-08 & 8.994e+04 & 2.670e-12\\
3.949e-01 & 3.188e-01 & 2.883e+00 & 8.628e-01 & 6.095e+01 & 2.688e-04 & 2.538e+03 & 1.124e-08 & 9.905e+04 & 2.139e-12\\
4.146e-01 & 3.542e-01 & 3.067e+00 & 7.741e-01 & 6.692e+01 & 2.041e-04 & 2.795e+03 & 8.909e-09 & 1.091e+05 & 1.714e-12\\
4.353e-01 & 3.930e-01 & 3.266e+00 & 6.885e-01 & 7.350e+01 & 1.548e-04 & 3.078e+03 & 7.067e-09 & 1.201e+05 & 1.374e-12\\
4.571e-01 & 4.349e-01 & 3.482e+00 & 6.072e-01 & 8.075e+01 & 1.174e-04 & 3.390e+03 & 5.609e-09 & 1.323e+05 & 1.101e-12\\
4.800e-01 & 4.797e-01 & 3.715e+00 & 5.311e-01 & 8.872e+01 & 8.889e-05 & 3.732e+03 & 4.454e-09 & 1.457e+05 & 8.825e-13\\
5.040e-01 & 5.273e-01 & 3.969e+00 & 4.606e-01 & 9.751e+01 & 6.728e-05 & 4.110e+03 & 3.539e-09 & 1.604e+05 & 7.073e-13\\
5.293e-01 & 5.779e-01 & 4.244e+00 & 3.962e-01 & 1.072e+02 & 5.089e-05 & 4.526e+03 & 2.813e-09 & 1.767e+05 & 5.672e-13\\
5.560e-01 & 6.317e-01 & 4.544e+00 & 3.382e-01 & 1.178e+02 & 3.848e-05 & 4.984e+03 & 2.237e-09 & 1.945e+05 & 4.487e-13
\enddata
	\tablenotetext{a}{Differential Intensity units: (m$^2$ s sr GV)$^{-1}$.}
\end{deluxetable}

\end{document}